\documentclass[twocolumn]{aa}
\usepackage{graphicx}
\usepackage[figuresright]{rotating}
\usepackage{longtable}
\begin{document}
%
   \title{New Herbig Ae/Be stars confirmed via high-resolution optical spectroscopy\thanks{Based on observations collected at the ESO-MPG 2.2 meter telescope at la Silla 
Observatory, Chile (program IDs: 072.A-9006, 073.A-9008, 079.A-9014, 081.A-9003)}}
 
 \titlerunning{New Herbig Ae/Be stars.}
 

   \author{A. Carmona \inst{1,}\thanks{This research was 
    		partly conducted while A. Carmona was at the
		    Max-Planck Institute for Astronomy in Heidelberg and 
		    the European Southern Observatory in Garching.}
          \and
          M.E. van den Ancker \inst{2}          
          \and
          M. Audard \inst{1}   
         \and
          Th. Henning \inst{3}
          \and
          J. Setiawan \inst{3} 
          \and
	      J. Rodmann \inst{4}
          }

   \offprints{A. Carmona\\
              \email{Andres.Carmona@unige.ch}}
             
   \institute{ISDC Data Centre for Astrophysics \& Geneva Observatory, University of Geneva, chemin d'Ecogia 16, 1290 Versoix, Switzerland
            \and
              European Southern Observatory, 
              Karl Schwarzschild Str 2 , 85748 Garching bei M\"unchen, Germany
            \and
	          Max-Planck Institute for Astronomy, K\"onigstuhl 17, 69117 Heidelberg, Germany
            \and
             ESA/ESTEC, Space Environments \& Effects (TEC-EES), 2200 AG Noordwijk, The Netherlands.
                }

   \date{}
   
   \abstract{
    We present FEROS high-resolution (R$\sim$ 45000) optical spectroscopy of 34 Herbig Ae/Be star candidates
    with previously unknown or poorly constrained spectral types. 
    A total of 32 sources are from the Th\'{e} et al. (1994) catalog and
    two are new nearby Herbig Ae/Be star candidates from Vieira et al. (2003).  
    Within the sample, 16 sources are positionally coincident with nearby (d$<$250 pc) star-forming regions  (SFRs). 
    All the candidates have reported infrared excess.
    We determine the spectral type and luminosity class of the sources, 
    derive their radial and projected rotational velocities,
    and constrain their distances employing spectroscopic parallaxes and photometry from the literature. 
    We confirm 13 sources as Herbig Ae/Be stars and find one classical T Tauri star.
	 Three sources are emission line early-type giants (B, A, and F stars with luminosity class III) and may be Herbig Ae/Be stars.
    One source is a main-sequence A-type star.
    Fourteen sources are post-main-sequence giant and supergiant stars (7 with H$\alpha$ emission and 7 without). 
    Two sources are extreme emission-line stars and no accurate spectral classification was possible because of strong veiling.
    Most of the sources appear to be background stars at distances over 700 pc.
    We show that high-resolution optical spectroscopy is a crucial tool for distinguishing young stars (in particular Herbig Be stars) from 
    post-main sequence stars in samples taken from emission-line star catalogs based on low-resolution spectroscopy.
    Within the sample, three young stars (CD-38 4380, Hen~3-1145, and HD~145718) and 
    one early-type luminosity class III giant with emission lines (Hen 3-416)
    are at distances closer than 300~pc and are positionally coincident with a nearby SFR.
    These 4 sources are likely to be nearby young stars and are interesting for 
    follow-up observations at high-angular resolution.
    Furthermore, seven confirmed Herbig Ae/Be stars at $d>700$~pc
    (Hen~2-80, Hen~3--1121~N\&S, HD 313571, MWC~953,  WRAY~15-1435, and Th~17-35) 
    are inside or close ($<5'$) to regions with extended 8 $\mu$m continuum emission and
    in their 20' vicinity have astronomical sources characteristic of SFRs 
    (e.g., HII regions, molecular clouds,
    dark nebulae, masers, young stellar-objects). These 7 sources are likely to be members of SFRs.
    These regions are attractive for future studies of their stellar content.
    
    }

   \maketitle
%
\begin{table*} 
\begin{minipage}[t]{\textwidth}
\scriptsize
\caption{Positions and extent in galactic coordinates of the star forming regions (SFR) employed.}
\centering                          
\renewcommand{\footnoterule}{}  
\begin{tabular}{l c c c c c c c l r}  
\hline\hline                 
          &                                                                                 & l1\footnote{The SFR is assumed to be a rectangle in the sky.
The galactic coordinates l1, l2,  b1, and b2 define such rectangle.} & l2 & b1 & b2 & d \\
Name &  Object \footnote{RN means reflection nebulae and NOBA means nearby OB association} &  [deg] &  [deg] &  [deg] &  [deg] &  [pc] &  References &  \\
\hline 
Sgr R1 &  RN &  7 &  14 &  -2 &  1 &  1560 &  van den Bergh (1966) &  \\
Vul R1 &  RN &  53 &  57 &  0 &  4 &  440 &  van den Bergh (1966) &  \\
Cyg OB4 &  NOBA &  81 &  85 &  -9 &  -6 &  1000 &  de Zeeuw et al. (1999) &  \\
Cyg OB7 &  NOBA &  84 &  96 &  -5 &  9 &  740 &  de Zeeuw et al. (1999) &  \\
Lac OB1 &  NOBA &  94 &  107 &  -19 &  -7 &  368 &  de Zeeuw et al. (1999) &  \\
Cep OB2 &  NOBA &  96 &  108 &  -1 &  12 &  615 &  de Zeeuw et al. (1999) &  \\
Cep OB6 &  NOBA &  100 &  110 &  -2 &  2 &  270 &  de Zeeuw et al. (1999) &  \\
Cep R2 &  RN &  103 &  112 &  10 &  15 &  700 &  van den Bergh (1966) &  \\
Cep R1 &  RN &  105 &  109 &  3 &  7 &  700 &  van den Bergh (1966) &  \\
Cep OB3 &  NOBA &  108 &  113 &  1 &  7 &  960 &  de Zeeuw et al. (1999) &  \\
Cep OB4 &  NOBA &  116 &  120 &  3 &  7 &  845 &  de Zeeuw et al. (1999) &  \\
Cam OB1 &  NOBA &  130 &  153 &  -3 &  8 &  900 &  de Zeeuw et al. (1999) &  \\
$\alpha$Persei &  NOBA &  140 &  155 &  -11 &  -3 &  177 &  de Zeeuw et al. (1999) &  \\
Per OB2 &  NOBA &  156 &  164 &  -22 &  -13 &  318 &  de Zeeuw et al. (1999) &  \\
Per R1 &  RN &  156 &  161 &  -22 &  -17 &  400 &  van den Bergh (1966) &  \\
Tau R1 &  RN &  166 &  167 &  -24 &  -23 &  125 &  van den Bergh (1966) &  \\
Tau R2 &  RN &  173 &  175 &  -17 &  -13 &  125 &  van den Bergh (1966) &  \\
Ori OB1 &  NOBA &  197 &  215 &  -26 &  -12 &  500 &  de Zeeuw et al. (1999) &  \\
Mon OB1 &  NOBA &  201 &  205 &  -3 &  3 &  715 &  de Zeeuw et al. (1999) &  \\
Mon R1 &  RN &  201 &  204 &  -1 &  3 &  715 &  van den Bergh (1966) &  \\
Ori R1 &  RN &  204 &  207 &  -18 &  -13 &  500 &  van den Bergh (1966) &  \\
Ori R2 &  RN &  208 &  213 &  -21 &  -18 &  500 &  van den Bergh (1966) &  \\
Cma R1 &  RN &  222 &  226 &  -4 &  -2 &  1315 &  van den Bergh (1966) &  \\
Col 121 &  NOBA &  222 &  244 &  -15 &  -3 &  592 &  de Zeeuw et al. (1999) &  \\
Vela OB2 &  NOBA &  255 &  270 &  -15 &  -2 &  410 &  de Zeeuw et al. (1999) &  \\
Trumpler 10 &  NOBA &  255 &  270 &  -2 &  4 &  366 &  de Zeeuw et al. (1999) &  \\
Sco OB2 \--5 &  NOBA &  273 &  292 &  -20 &  5 &  145 &  de Zeeuw et al. (1999) &  \\
Low.Cen.Crux &  NOBA &  285 &  312 &  -10 &  21 &  118 &  de Zeeuw et al. (1999) &  \\
Sco OB2\--4 &  NOBA &  292 &  313 &  -10 &  16 &  145 &  de Zeeuw et al. (1999) &  \\
Up.Cen.Lup &  NOBA &  310 &  345 &  0 &  25 &  140 &  de Zeeuw et al. (1999) &  \\
Sco OB2\--3 &  NOBA &  313 &  337 &  5 &  31 &  145 &  de Zeeuw et al. (1999) &  \\
Sco OB2\--1 &  NOBA &  330 &  3 &  -19 &  7 &  145 &  de Zeeuw et al. (1999) &  \\
Sco OB2\--2 &  NOBA &  337 &  3 &  7 &  32 &  145 &  de Zeeuw et al. (1999) &  \\
Sco R1 &  RN &  346 &  2 &  13 &  23 &  160 &  van den Bergh (1966) &  \\
\hline 
\end{tabular} 
\end{minipage}
\end{table*} 
%
\section{Introduction}
Herbig Ae/Be stars are intermediate-mass (2-8 M$_\odot$)
pre-main sequence (PMS) stars.
They exhibit emission lines (e.g., H$\alpha$, H$\beta$, \ion{Ca}{ii})
in their optical spectra
and infrared (IR) excess in their spectral energy distributions.
These observational characteristics provide indirect evidence 
that Herbig Ae/Be stars have an accreting circumstellar disk.
The infrared excess is interpreted as emission from small dust grains present
in the hot surface layer of the disk.
By analogy  with the lower-mass T Tauri stars (e.g., Hartmann 1999, Muzerolle et al. 2004),
the \ion{H}{i} emission lines can be interpreted as originating in the magnetospheric 
accretion shock\footnote{The origin of the \ion{H}{i} emission lines in Herbig Ae/Be stars is in fact controversial. 
Several authors have alternative scenarios for magnetospheric accretion (e.g., strong stellar winds, outflows, direct disk accretion)  
to explain the origin of the \ion{H}{i} lines (e.g., B\"ohm \& Catala 1993,  Mottram et al. 2007).  
The magnetospheric accretion model can explain the H$\alpha$ line and its spectropolarimetry signal 
in Herbig Ae stars (e.g., Pontefract et al. 2000, Vink et al. 2002, 2005).
However, this is less clear in the case of Herbig Be stars, 
since wind or outflows contributions to the H$\alpha$ line are likely, 
and the spectropolarimetry signal does not unambiguously support the magnetospheric accretion scenario (Mottram et al. 2007).    
As several stars in our sample are Herbig Be stars, it could well be that different emission mechanisms work for different stars in our sample.}
when the gas of the disk reaches the surface of the star with free-fall velocities of a few hundred km/s.
More recently, spatially resolved dust and molecular line observations
in the millimeter and sub-millimeter domain (e.g., Mannings and Sargent 1997, Semenov et al. 2005),
together with scattered light coronographic imaging (e.g., Fukagawa et al. 2004, Grady et al. 2005) 
provided direct evidence that Herbig Ae/Be stars are effectively surrounded by a disk
(for a detailed review about Herbig Ae/Be stars see Waters and Waelkens 1998).

From the observational point of view, 
bright nearby (d$<$250 pc)  Herbig Ae/Be stars  (and CTTS) are particularly relevant,
because they permit detailed studies of the structure of their disks. 
Disks are interesting because they play a key role in early stellar evolution and 
are the sites of planet formation. 
In nearby sources, the disk can be spatially resolved with 8 -- 10 m class telescopes and infrared and mm interferometers.
In bright sources, high--resolution spectroscopy in the near and mid-IR can be obtained to study the
gas in the disk (see reviews by Najita et al. 2007 and Carmona 2010).
Since the amount of identified nearby PMS
with spatially resolved disks is still relatively small,
the identification of bright nearby Herbig Ae/Be stars
is an important step for future observational studies of protoplanetary disks.

%
\begin{table*} 
\begin{minipage}[t]{\textwidth}
\caption{Studied stars, SFRs positionally coincident and summary of the observations.}
\label{table:1}      
\centering   
\scriptsize                      
\renewcommand{\footnoterule}{}  
\begin{tabular}{l c c c c c c c c c c }        
\hline\hline                 	    
&  $\alpha$ (J2000.0)  & $\delta$ (J2000.0)  & l & b    &  & d$_{\rm SFR}$ &  t$_{exp}$ & &  Date(s)\\
Star\footnote{All sources from Table IVb of Th\'e et al. (1994), except for Hen 3-1145 (Table II), Hen 2-80 (Table IVa) and
CD-38 4380 and HD145718 (Vieira et al. 2003).} & [~h~~m~~~ s~ ] & [~$^\circ$ ~~'~~~ ''~ ]& [deg] & [deg] & SFR & [pc] & [s] & S/N\footnote{The S/N is the average S/N in the continuum close to the H$\alpha$ line.} &  [yyyy-mm-dd]  \\
\hline 
CD-38 4380       & 08 23 11.86 & -39 07 01.5	  &257.32	& -1.06 &Gum Nebula                             & 200 \-- 240 & 1200 & 25 & 2004-04-03 \\
WRAY 15-488    & 10 01 48.11 & -59 12 12.5    & 282.70 & -3.17  & ScoOB2\--5                             & 145            & ~800 & 25 & 2004-04-04 \\
WRAY 15-522    & 10 12 12.42 & -62 32 33.1    & 285.69 & -5.13  & ScoOB2\--5                             & 145            & 1500 & 25 & 2004-04-05 \\
Th 35-41             & 10 25 40.07 & -58 22 17.4    & 284.78 & -0.73  & ScoOB2\--5                             & 145            & 1800 & 15 & 2004-04-08 \\
Hen 3-416          & 10 25 44.51 & -58 33 52.2    & 284.89 & -0.89  & ScoOB2\--5                             & 145             & 1200 & 30 & 2004-04-07 \\
WRAY 15-566    & 10 25 51.36 & -60 53 13.2    & 286.13 & -2.86  & Low.Cen.Crux \-- ScoOB2\--5 & 118 \-- 145  & 1800  & 10 & 2004-04-07 \\
HD 305773       & 10 56 03.88 & -60 29 37.6    & 289.18 & -0.74  & Low.Cen.Crux \-- ScoOB2\--5 & 118 \-- 145  & ~700 & 100 &2008-05-21 \\
WRAY 15-770    & 11 11 28.49 & -63 00 23.7    & 291.87 & - 2.30 & Low.Cen.Crux \-- ScoOB2\--5  & 118 \-- 145  & 1200 & 10 & 2004-04-05 \\
Hen 2-80            & 12 22 23.18 & -63 17 16.8    & 299.67 &  -0.60 & Low.Cen.Crux \-- ScoOB2\--4  & 118 \-- 145  & 1800 & 25 & 2004-04-08 \\
Hen 3-823          & 12 48 42.39 & -59 54 35.0    & 302.59 &  2.96 & Low.Cen.Crux \-- ScoOB2\--4  & 118 \-- 145  & ~900 & 60 & 2008-04-21 \\
Th 17-35             & 13 20 03.59 & -62 23 54.0    & 306.24 &  0.29 & Low.Cen.Crux \-- ScoOB2\--4  & 118 \-- 145 & 1800 & 25 & 2004-04-08 \\
WRAY 15-1104   & 13 29 51.02 & -56 06 53.7   & 308.30 &  6.36 & Low.Cen.Crux \-- ScoOB2\--4  & 118 \-- 145 & 1500 & 40 & 2004-04-06 \\
WRAY 15-1372 
                            & 15 53 50.59 & -51 43 05.1    & 329.21   &1.59  & Up.Cen.Lup                            & 140              & 1500 & 30 & 2004-04-08 \\
Hen 3-1121N    & 15 58 09.62 & -53 51 18.3 & 328.345 & -0.463 & ... & - & 1500 & 70 & 2008-04-21 \\
Hen 3-1121S    & 15 58 09.67 & -53 51 34.9 & 328.342 & -0.466 & ... & - & 1500 & 85 & 2008-04-24 \\                    
Hen 3-1145     & 16 08 54.69 & -39 37 43.1    & 339.21 & 8.95    & Up.Cen.Lup   & 140 & 2000 & 16& 2004-04-08 \\
WRAY 15-1435   & 16 13 06.68 & -50 23 20.0 & 332.36  &  0.58  & ... & - & 1800 & 55 & 2008-04-24 \\
HD 145718      & 16 13 11.59 & -22 29 06.6    & 352.43 & 20.44    & $\rho$Oph \-- Sco OB2 & 110 \-- 160& 1000 & 20 & 2004-04-08 \\
HD 152291       & 16 54 24.20 & -40 39 09.0 & 344.40  &  1.88  & Up.Cen.Lup   & 140 & ~600 &  100& 2008-04-21 \\
Hen 3-1347     & 17 10 24.15 & -18 49 00.7    & ~~~4.10 & 12.26   & ... & - & 1800 & 70 & 2007-08-04 \\
WRAY 15-1651   & 17 14 45.03 & -36 18 38.4  & 350.27	&  1.35 & ... & - & 2400 & 17 & 2008-05-01 \\
WRAY 15-1650   & 17 15 32.79 & -55 54 22.7    & 334.28  &	-10.07  & ... & - & 900  & 8& 2007-08-17 \\  
HD 323154        & 17 23 02.36 &	-39 03 52.5    & 348.96  & -1.57   & ... & - & 600  & 120 &  2007-08-04 \\     	
WRAY 15-1702   & 17 24 30.88 & -37 34 27.7  & 350.35	& -0.97 & ... & - & 1700 &8 &  2008-04-30 \\
MWC 878        & 17 24 44.70 &	-38 43 51.4	    & 349.42  &	-1.65   & ... & - & 900 & 70 & 2007-08-04 \\ 
AS 231
         & 17 30 21.66 &	-33 45 29.6	    & 354.18  & 0.17	& ... & - & 1400 & 140 & 2004-07-31 \\ 
Hen 3-1428      & 17 35 02.49 & -49 26 26.4    & 341.41 &~-9.03& ... & - & 1200 & 50 & 2007-08-07 \\
HD 320156     & 17 37 58.51 &	-35 23 04.3     & 353.66  & -2.03 & ... & - & 600 & 90 & 2007-08-07 \\
MWC 593        & 17 49 10.16 &	-24 14 21.2	    & ~~~4.43 & ~~1.75 & ... & - & 720 & 85 & 2007-07-31 \\
HD 313571        & 18 01 07.18 & -22 15 04.0		& ~~~7.53 &	~~0.39 & ... & - & 900 & 70 & 2007-08-02 \\
MWC 930        & 18 26 25.24 & -07 13 17.8		& ~23.65  &	~~2.23 & ... & - & 2100 & 50 & 2007-08-04 \\
MWC 953        & 18 43 28.43 & -03 46 16.9    	& ~28.69  & ~~0.05 & ... & - & 1200 & 60 & 2007-07-31 \\
AS 321         & 18 47 04.80 &	-11 41 02.3	    & ~22.02  & ~-4.35 & ... & - & 1500 & 75 & 2004--07-31\\
MWC 314        & 19 21 33.97 & +14 52 57.0	    & ~49.57  & ~~0.25 & ... & - & 900 & 65 & 2007-08-04 \\
\hline
\end{tabular}
\end{minipage}
\end{table*}

Herbig Ae/Be stars were initially identified based on the presence of emission lines (i.e. H$\alpha$) in their optical spectra
and their physical association with a dark cloud or nebulosity (e.g., Herbig, 1960).
Herbig (1960) used the last condition to exclude the post-main sequence B[e] stars (i.e. giant or supergiant B-type stars with emission lines).
Thanks to the advent of IR space observatories such as IRAS, ISO, and Spitzer 
this last criterion has been relaxed and replaced by the presence of  near- or far-IR excess, in addition to the emission lines,
as membership criteria to the Herbig Ae/Be stellar group (e.g., Finkenzeller \& Mundt 1984; Th\'e et al. 1994; Vieira et al. 2003).

However, 
published samples of Herbig Ae/Be stars may well be contaminated with other 
classes of objects. One should bear in mind three aspects of the identification: 
({\it i\,}) in general,  
 Herbig Ae/Be star candidates have been identified in surveys for emission-line stars based on low-resolution data, in particular slit-less spectra
(this makes no difference on the detection of the emission lines, 
but it matters for the determination of the luminosity class, see below);
({\it ii\,})  post-main sequence B[e] supergiants can also have IR excess
(e.g., Miroshnichenko et al. 2005 and references therein);
and ({\it iii\,}) as beam sizes for infrared observations employed in previous studies have 
typically been large (e.g., 30" in the case of IRAS), confusion with other 
infrared sources may have occurred.

Since hydrogen lines are observed in emission in Herbig Ae/Be stars, 
the hydrogen lines width, 
the usual means for determining the luminosity class, cannot be used.
Thus the observation of gravity sensitive lines (e.g., \ion{N}{ii} at 3995~\AA,  
\ion{Si}{ii} at 4128 and 4131~\AA,  \ion{C}{ii} at 4267~\AA, \ion{Si}{iii}  at 4553 and 4561~\AA,
and \ion{O}{ii} lines at 4070 and 4976~\AA) is required.
However, these lines are relatively weak and are barely visible in low-resolution spectra.
Therefore, low spectral resolution studies that have identified Herbig Ae/Be candidates 
have the important limitation that background B[e] supergiants
can be mistakenly classified as Herbig Ae/Be stars.
Consequently, to confirm that a Herbig Ae/Be candidate is indeed a young star - and not a post-main sequence object -
observations at high spectral resolution are necessary. 

In this paper, we present the results of a high-resolution optical spectroscopy 
campaign aimed at identifying and characterizing new nearby Herbig Ae/Be stars. 
We obtained FEROS high-resolution (R$\sim$45000) optical spectra of 34 candidates to Herbig Ae/Be  stars.
We studied sources positionally coincident with 
nearby (d$<$250 pc) star-forming regions (SFRs) and ``isolated" sources. 
Our goal was to determine whether the candidates belong to the Herbig Ae/Be stellar group 
by searching emission lines in their spectra and by determining their spectral type and luminosity class.
We then constrained their distances employing spectroscopic parallaxes
and derived their radial and projected rotational velocities.
In the case of sources positionally coincident with SFRs, 
we used the estimated distance to determine whether the sources are members of nearby SFRs.  
Finally, for the confirmed Herbig Ae/Be stars that are not members of nearby SFRs,
we searched the Spitzer archive for 8 $\mu$m imaging and the SIMBAD database to find evidence for extended near-infrared emission
and astronomical objects characteristic of SFRs 
(e.g., HII regions, molecular clouds, dark nebulae, masers, young stellar-objects).
Our aim was to find evidence of whether these distant Herbig Ae/Be stars might be members of distant SFRs.

\begin{figure*}
\centering
\includegraphics[angle=0,width=\textwidth]{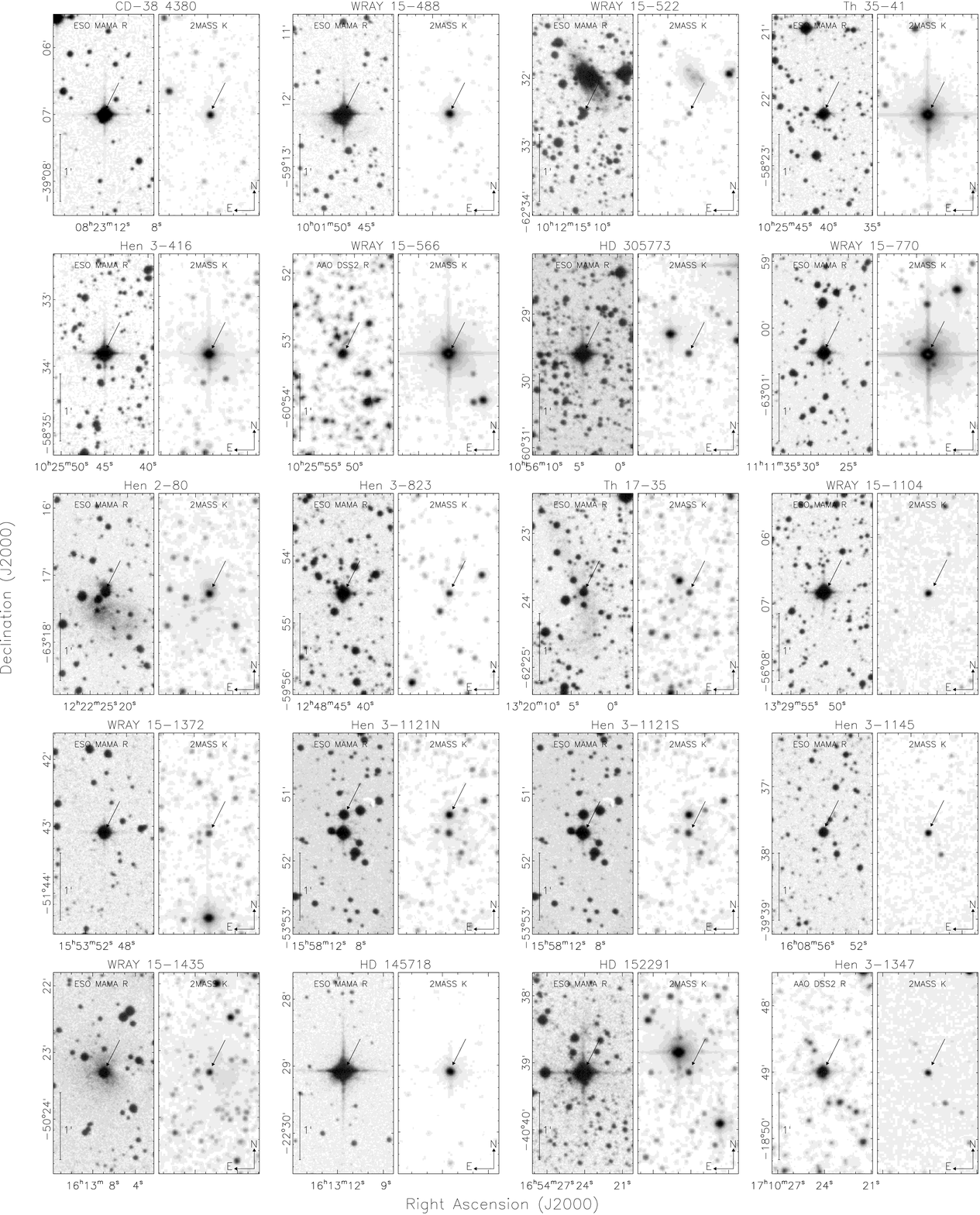}
    \caption{Optical (R band) and near-IR (K band) images of the 0.5' $\times$ 1' field
centered on the target stars. The target is indicated by an arrow.}
\end{figure*}

\begin{figure*}
\centering
\includegraphics[angle=0,width=\textwidth]{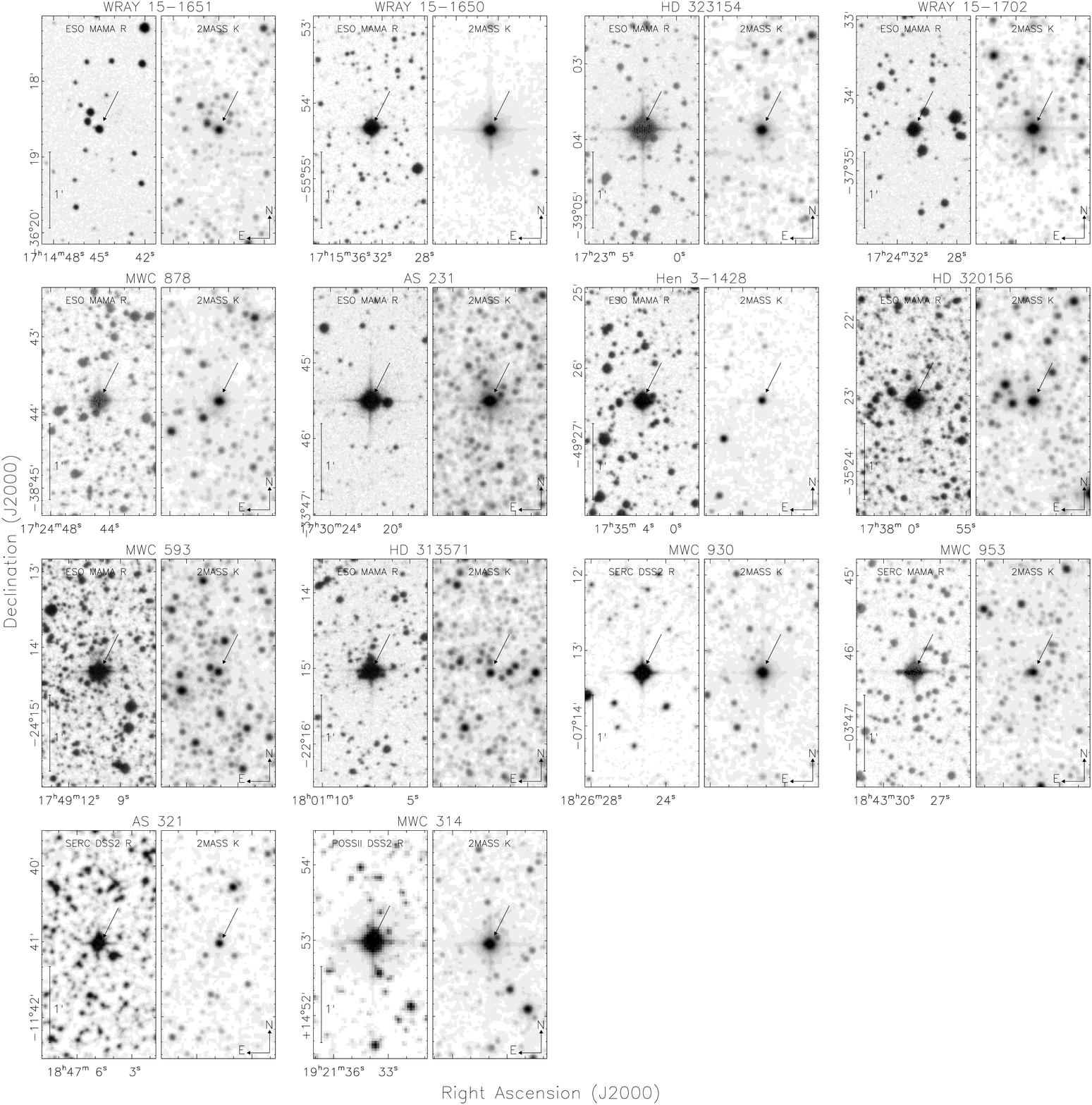}
    \caption{Optical (R band) and near-IR (K band) images of the 0.5' $\times$ 1' field
centered on the target stars. The target is indicated by an arrow.}
\end{figure*}

\section{Observations}
\subsection{Target Selection}
We selected the sources from the catalog of Th\'{e} et al. (1994).
This catalog is divided in six tables:
Table I presents the Herbig Ae/Be members and candidate members.
Table II, F-type stars potential candidates.
Table III, extreme emission line objects.
Table IVa/b, other early type emission line stars with IR excess from Allen \& Swings (1976) 
and Dong \& Hu (1991) respectively.
Table V, non-emission line early type shell stars and young stellar candidates.
Table VI, emission line stars rejected as Herbig Ae/Be candidates.
We concentrated only on sources  from Tables I, II, III, and IVa/b that are observable
from the Southern Hemisphere.
First, we selected the sources of unknown or poorly constrained spectral types
that are positionally coincident with nearby (d$<$ 250 pc) SFRs.
Second, we selected isolated Herbig Ae/Be stars candidates of 
unknown or poorly constrained spectral type that 
exhibited $V$ band magnitudes brighter than 14 (i.e. to obtain a good quality 
spectrum in a reasonable amount of exposure time).
To establish the association of a Herbig Ae/Be star candidate to a nearby SFR,
we assumed a projected rectangular geometry in the sky for the SFR.
We associated the candidate Herbig Ae/Be star  to a SFR 
if it is positionally within the sky region covered by the SFR.
We employed the SFRs from   
the ``{\it Study of Reflection Nebulae}" by van den Bergh (1966),
the ``{\it Hipparcos census of nearby OB associations}" by de Zeeuw et al. (1999),
and the distance measurements of Lynds galactic dark nebulae by Hilton \& Lahulla (1995).
In Table 1, we show the positions and extent of the SFRs employed.

The sample was complemented with two new Herbig Ae/Be star candidates (CD-38 4380 and HD 145718) 
from Vieira et al (2003) associated with a SFR at d$<$250 pc.
In Table 2, we present a summary of the studied stars, their coordinates 
and the positionally coincident SFR.
The coordinates used for the Herbig Ae/Be star candidates
are the coordinates of the brightest K band 2MASS object in the 30'' 
vicinity of the coordinates given by the Th\'{e} et al. catalog.
In Figures 1 and 2,  we present optical R band (ESO-MAMA) and K band (2MASS) images of the 0.5' $\times$ 1' 
field centered on the target position\footnote{Images were obtained using the Aladin software from the CDS, Strasbourg: 
http://aladin.u-strasbg.fr}.

\subsection{Observations \& Data Reduction} 
High-resolution optical spectra of the sources were obtained 
in  April-July 2004, July-August 2007, and May 2008, 
using the Fiber-fed  Extended Range Optical Spectrograph (FEROS; Kaufer et al. 1999) at the ESO/MPG 2.2 meter telescope in la Silla Observatory\footnote{http://www.ls.eso.org/lasilla/sciops/2p2/E2p2M/FEROS}.
FEROS covers the complete optical spectral region (3500 \-- 9200~\AA) in one exposure with a resolution of R $\sim$ 45000.
One of the two fibers was positioned at the location of the target star,
the second fiber was positioned at the sky.
Calibration flat fields, darks frames and Th-Ar lamp calibration spectra were
observed at the beginning of the night for each set of observations.
The data were reduced using the FEROS-DRS pipeline provided by ESO.
The typical achieved signal-to-noise ratio (S/N) is over 30. 
In some cases the S/N achieved is of the order of 60 or larger, 
only in a few cases was the S/N under 30 (see Table 2).

\begin{figure*}
 \centering
 \includegraphics[angle=90,width=\textwidth]{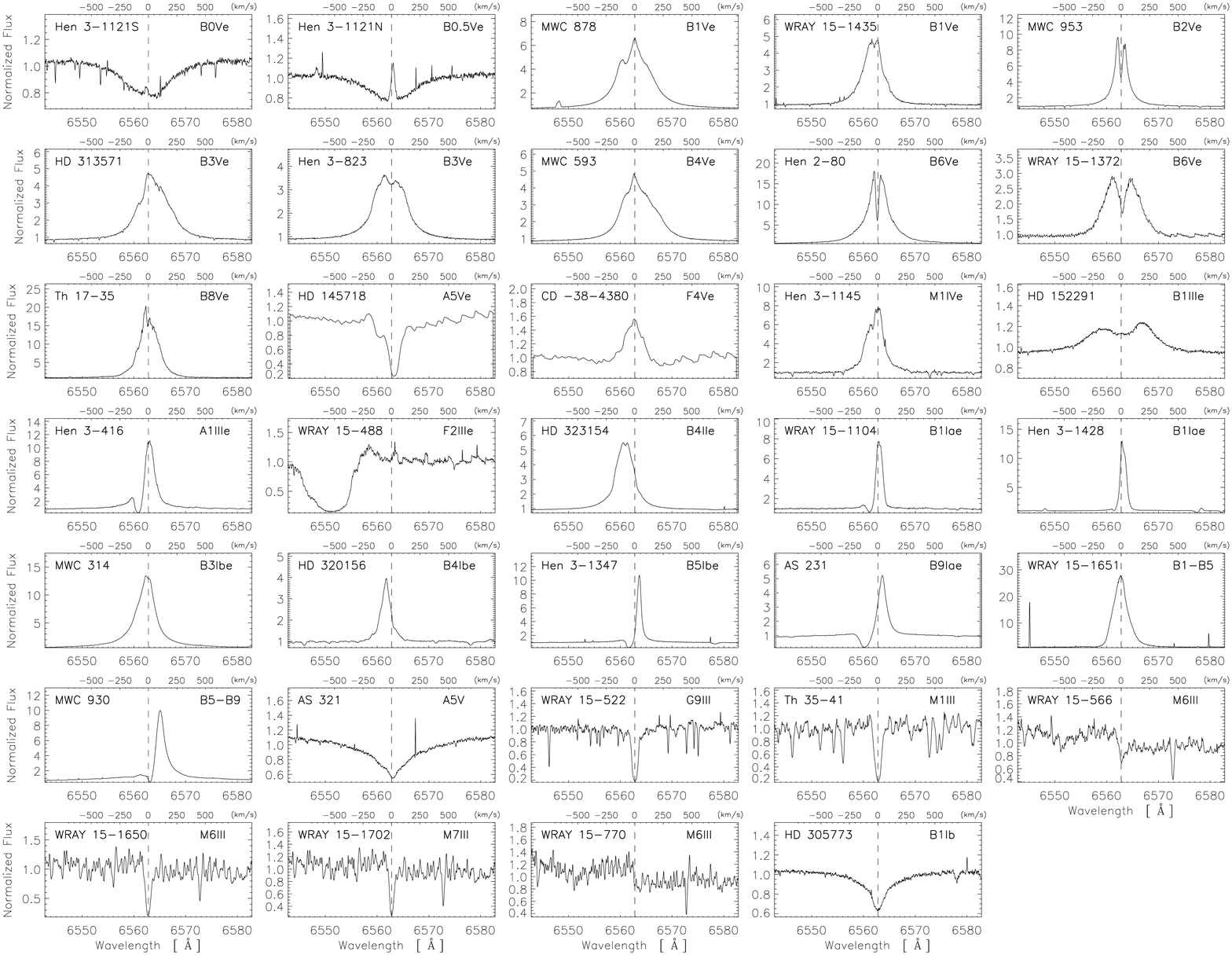}
      \caption{Optical spectra of our sources centered on the H$\alpha$ line. 
      The spectra are barycentric and radial--velocity corrected (see Table 3).  
      A velocity scale is given on top of each panel.}
         \label{FigHalpha}
   \end{figure*}
   
\begin{figure*}
   \centering                  
      \includegraphics[width=0.7\textwidth]{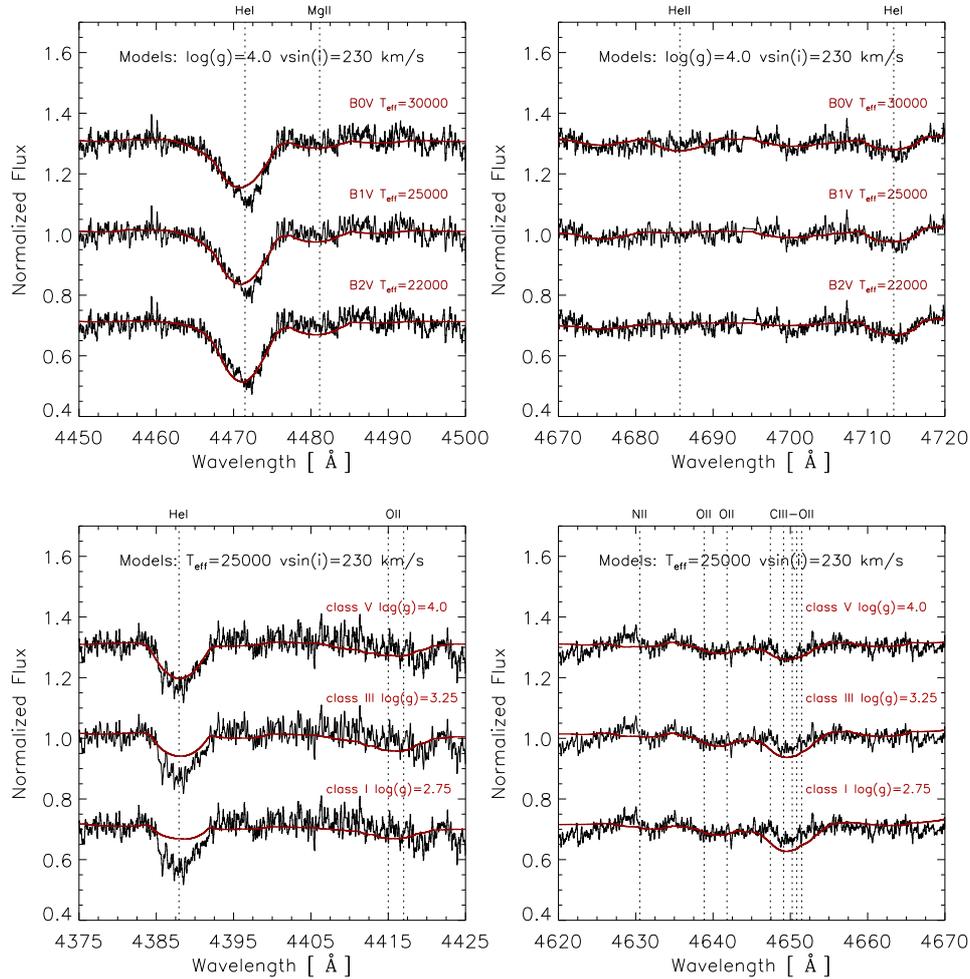}
   \caption{  Example of FEROS spectra of the Herbig Be star WRAY 15-1435 (spectral type B1Ve), 
   together with rotationally broadened TLUSTY models (in red). The spectra are shown at the rest velocity of the star.
    See text for a description.}
            \label{fig:}
   \end{figure*}

 \begin{figure*}
   \centering                  
   \includegraphics[width=0.7\textwidth]{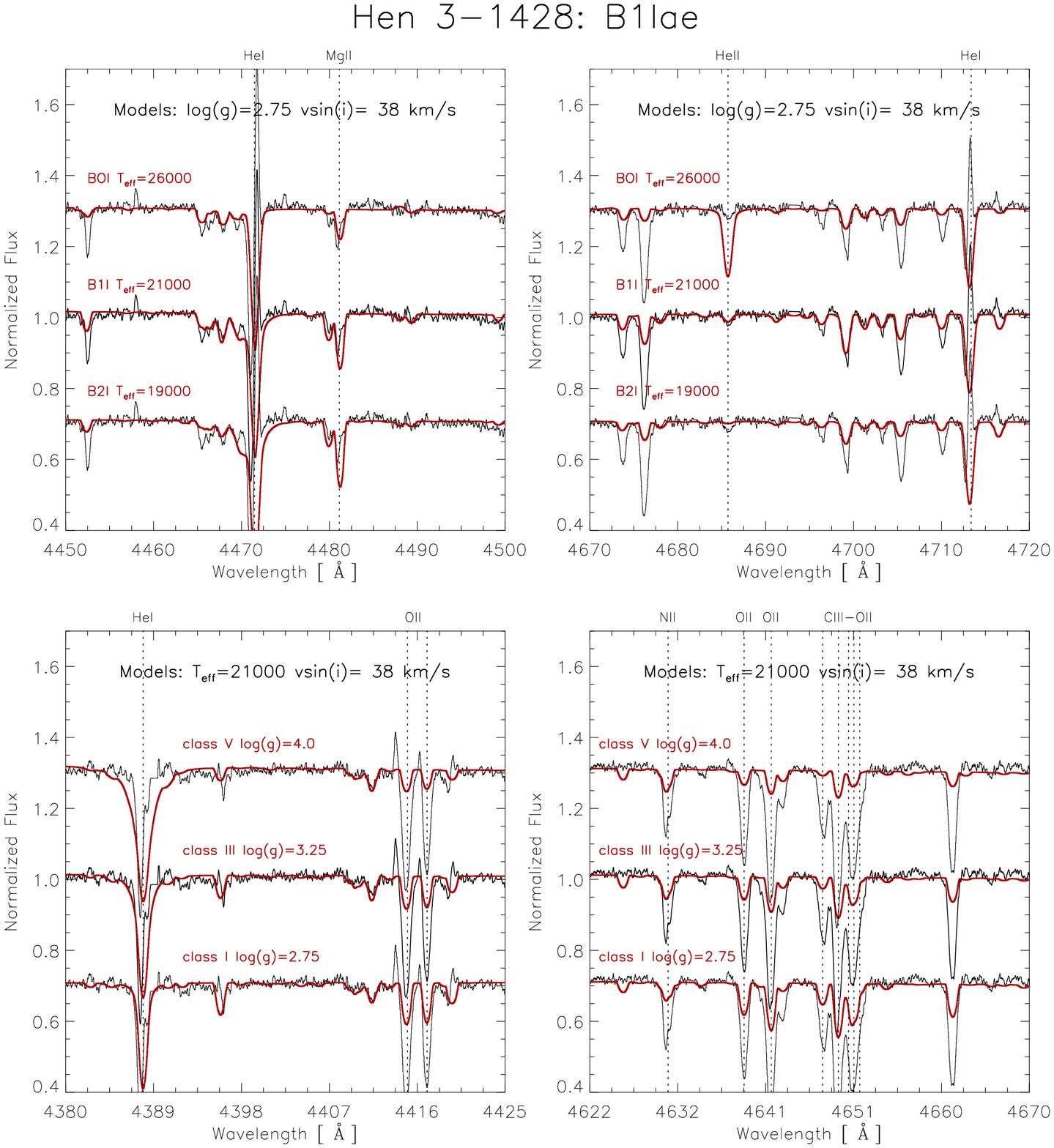}
   \caption{  Similar to Fig. 4 for the supergiant star HD 3-1428 (spectral type B1Iae). 
   The \ion{He}{i} line at 4471 \AA\ displays a P-Cygni profile. 
    }
            \label{fig:}
   \end{figure*}

\section{Analysis}
\subsection{Spectral Classification}
The spectral type and the luminosity class have been determined by careful comparison with 
the high-resolution (R$\sim$80000) spectra from the UVES  
Paranal Observatory Project spectral library\footnote{http://www.sc.eso.org/santiago/uvespop/} 
(Bagnulo et al. 2003), 
the ELODIE (R$\sim$42000) spectral library\footnote{http://atlas.obs-hp.fr/elodie/} (Soubiran et al. 1998),
the BLUERED\footnote{http://www.inaoep.mx/~modelos/bluered/bluered.html}  
high-resolution (R$\sim$500000) synthetic spectral library (Bertone et al. 2008), 
and in the particular case of early B-type stars\footnote{TLUSTY models describe better than the BLUERED models the observed strength of the \ion{He}{i} lines in early B-type stars.}, non-LTE TLUSTY (R$\sim$300000) models of B-type stars (Lanz \& Hubeny 2007).

A dedicated interactive IDL-based software was developed\footnote{Available from A. Carmona.} 
to visualize the normalized target spectrum simultaneously (i.e. over-plotted) with a normalized template spectrum at 
a user selected spectral feature.
Target and template spectra were normalized either by the median or by a polynomial fit of order 2 
to the flux in the 20-100~\AA\ windows used for the spectral comparison.
Template spectra were degraded down to the FEROS resolution prior to the analysis.
The comparison software allows the target or template spectrum to be shifted in wavelength such that
target and template spectra had the absorption features at the
same wavelength (i.e. to correct for the radial velocity difference of the star and the template).
It displays the residuals in real time between the target and template spectrum 
(square root of the summed square of the difference between the normalized template 
and target spectrum), it permits calculation of equivalent widths, and it performs Gaussian fits to spectral lines.
In addition, the tool allows the effect of stellar rotation onto a spectral template to be visualized.
This refinement was necessary to deduce the spectral type and $v$ sin($i$) of fast rotating candidates.
For this we convolved the spectral template with the 1-d normalized kernel of
a rotationally broadened line.

We compared our FEROS spectra with the spectral templates,
spectral feature by spectral feature
by employing a window of width 20-100~\AA\ around the central wavelength of each feature.
The selected comparison spectral features were extracted primarily from the {\it``Atlas of stellar spectra"} by  Ginestet et al. (1992) 
and the recommended features by Morgan, Keenan and Kellman (1943) from {\it``An atlas of
stellar spectra with an outline of spectral classification"}.
Essentially, the procedure consisted of employing the Ginestet et al. and Morgan et al.
works to guide us in which spectral diagnostics to analyze or look at
and using the UVES, ELODIE, BLUERED, or TLUSTY spectral library templates 
to compare the presence or absence, and the strength and shape of absorption lines 
to our spectra.

First, the presence or absence of certain photospheric features allowed us to constrain, 
in a relatively straightforward way, the spectral type up to two or three spectral subclasses. 
The absence of a feature sets a lower or upper limit to the spectral range, and 
the presence of other spectral feature(s) sets the complementary lower or upper limit to narrow the possible spectral range.
Afterwards, ratios between lines (e.g., \ion{He}{i} at 4471 \AA\ and \ion{Mg}{ii} at 4481 \AA) provide an additional constraint to the spectral type range.
The key to the method is to use the absence of spectral lines as an additional spectral classification criterium 
to the strength of the observed absorption lines. 
Finally, a closer comparison with the spectral library templates allowed us to narrow the classification to one spectral sub-type. 
Several spectral lines were used to classify each source. 
In Appendix 1, we describe the diagnostics that we employed for each star.

Once the spectral type was found, we determined the luminosity class.
For this, we mainly relied on the diagnostics described in Ginestet et al. (1992),
specifically those shown in the plates describing the effect of the luminosity for each subspectral type.
In the case of M-type stars, we based the luminosity classification on the diagnostics proposed by Montes et al. (1999).
In general, surface gravity diagnostics are additional spectral features to the ones used for the spectral classification.
In cases where a few spectral types matched the spectra under study (e.g., in objects with large $v$ sin$(i)$), 
the template spectrum that exhibited the smallest residuals was
adopted as the best match for the spectrum of the target.
In Appendix 1, we describe in detail the spectral and luminosity class classification for each star individually.
In Col. 2 of Table 3, we summarize the results of the spectral classification of our sample.

In Figures 4 and 5, 
we present two examples of the spectral classification results.
We show the spectra of the Herbig Be star WRAY 14-1435 and the 
B-type supergiant star Hen 3-418, together with rotationally-broadened TLUSTY models.
Each panel presents the FEROS spectrum of the star with three TLUSTY models.
In the two panels at the top, 
we keep the surface gravity constant (i.e. luminosity class) and we vary T$_{\rm eff}$.
The theoretical spectrum in the middle corresponds to a TLUSTY model with T$_{\rm eff}$ of the spectral type and luminosity class found (see Table 3).
The upper and lower theoretical spectra display templates that are one spectral subclass distant in effective temperature.
These panels cover wavelengths including the \ion{He}{i} line at 4471 \AA, the \ion{Mg}{ii} line at 4481 \AA,
the \ion{He}{ii} line at 4686 \AA, and the \ion{He}{i} line at 4388 \AA,  
all important diagnostics for determining the spectral type of B-type stars.
In the two panels at the bottom, 
we keep the T$_{\rm eff}$ constant at a value corresponding to the spectral type found 
and vary the luminosity class of the theoretical spectrum from a dwarf (class V), to a giant (class III), and to a supergiant (class I).
These panels cover wavelengths including the \ion{N}{ii} line at 4630 \AA,
the \ion{O}{ii} lines at 4638 and 4641 \AA,  and the \ion{C}{iii} -  \ion{O}{ii} lines at 4650 \AA,
important diagnostics for establishing the luminosity class (i.e. log($g$)).
The typical uncertainty in the spectral classification is one spectral subclass and 0.5 in log(g) depending 
on the spectral-type and $v$ sin($i$).
\begin{table*} 
\scriptsize	
\begin{minipage}[t]{\textwidth}
\caption{Established spectral type, literature photometry, estimated distance, radial and projected rotational velocity, 
observed H$\alpha$ and forbidden line emission.}
\label{table:3}      
\centering                          
\begin{tabular}{l	c	c		c	c	c	c	c	c	c	c c c c c c }
\hline\hline                
       &          &                 T$_{\rm eff}$ &  $B$   &  $V$   & $E(B-V)$ & $A_V$ & $d_{\rm phot}$ &   $v ~{\rm sin}(i)$  & $v_r$  &     \\
Star  &   Sp.Type  &     [K]            & [mag] & [mag] & [mag]    & [mag]  & [kpc]                 &  [km/s]                   &  [km/s]  & H$\alpha$  & F.L.       \\
(1)   &  (2)   &  (3)           &     (4)            &   (5)    &    (6)  &  (7)       & (8)           &  (9)      &   (10)                & (11) & (12) \\[2mm]
\hline
\multicolumn{12}{c}{\it Confirmed Herbig Ae/Be and T Tauri stars}\\
\hline
Hen 3-1121S &   B0Ve     &  30000  &  11.6$\pm0.1$  & 11.1$\pm0.1$ & 0.85$\pm0.14$  &  2.7$\pm0.5$  &  3.1$^{+1.0}_{-0.8}$ 
& 240$\pm20$ & -28$\pm5$ &    I & [\ion{O}{i}]\\
Hen 3-1121N &   B0.5Ve &  27500  &     ...                  & 13.0$\pm0.4$ & 1.3$\pm0.7^\star$  &  4.1$\pm2.3$  &  3.2$^{+6.3}_{-2.1}$  
& 245$\pm20$ & -58$\pm5$ &    I & [\ion{O}{i}] [\ion{N}{ii}] \\
MWC 878 &   B1Ve &  25400  &        11.2$\pm0.1$  & 10.8$\pm0.1$  & 0.74$\pm0.15$  &  2.3$\pm0.5$  &  2.1$^{+0.9}_{-0.6}$   
&  ... & -11$\pm3^*$ &  I &     [\ion{O}{i}] [\ion{N}{ii}]\\
WRAY 15-1435 &   B1Ve &  25400  &        12.8$\pm0.5$  & 12.0$\pm0.5$  & 1.0$\pm0.7$  &  3.2$\pm2.2$  &  2.6$^{+5}_{-1.7}$  
& 230$\pm20$ & -30$\pm10$ &  I & [\ion{O}{i}] \\
MWC 953 &   B2Ve &  22000  &        12.0$\pm0.2$  & 10.9$\pm0.1$  & 1.3$\pm0.2$ &  4.0$\pm0.7$  & 0.8$^{+0.5}_{-0.3}$  
& $<7$         &  23$\pm2$ &  II &    [\ion{O}{i}] \\
HD 313571 &  B3Ve &  18700  & 10.3$\pm0.1$  & 9.9$\pm0.1$   &  0.57$\pm0.15$  &  1.8$\pm0.5$  &  0.9$^{+0.4}_{-0.3}$  
& 380$\pm20$ &  -50$\pm20$ & I &      [\ion{O}{i}] \\
Hen 3-823 &   B3Ve &  18700  & 10.5$\pm0.1$  & 10.4$\pm0.1$  & 0.33$\pm0.15$  &  1.0$\pm0.5$  &  1.6$^{+0.8}_{-0.5}$   
& 310$\pm20$ & -10$\pm10$ &   II &   [\ion{O}{i}] \\
MWC 593 &   B4Ve &  17000  & 10.5$\pm0.1$  & 10.0$\pm0.1$ & 0.65$\pm0.15$  &  2.1$\pm0.5$  &  0.7$^{+0.2}_{-0.2}$  
& 330$\pm30$ & -20$\pm10$ &   I &      [\ion{O}{i}] \\
Hen 2-80 &    B6Ve &  14000  & 14.0$\pm0.1$  & 13.2$\pm0.1$  & 0.97$\pm0.15$  &  3.0$\pm0.5$   &  1.6$^{+0.5}_{-0.4}$  & 
60$\pm$10 &  -22$\pm3$ & II &    [\ion{O}{i}][\ion{S}{ii}] \\
WRAY 15-1372 &  B6Ve &  14000  &  11.7$\pm0.1$  & 11.1$\pm0.1$  & 0.80$\pm0.15$  &  2.5$\pm0.5$  &  0.8$^{+0.2}_{-0.17}$  
& 280$\pm20$& -30$\pm$10 &  II &     ... \\
Th 17-35&    B8Ve &  11900  &   13.9$\pm0.1$  & 13.5$\pm0.1$ & 0.58$\pm0.15$  &  1.8$\pm0.5$  &  2.4$^{+0.8}_{-0.6}$  
& 210$\pm20$& -10$\pm5$ &  I &  [\ion{O}{i}] \\
HD 145718 &  A5Ve &  8200  & 9.3$\pm0.1$  &  8.9 $\pm0.1$ &  0.24$\pm0.15$  &  0.8$\pm0.5$  &  0.17$^{+0.05}_{-0.04}$ 
& 100$\pm10$ 		 & 0$\pm3$ &   IV & ... \\
CD-38 4380 &   F4Ve &  6590  & 10.4$\pm0.1$  & 10.0$\pm0.1$  & 0.0$\pm0.15$  &  0.0$^{+0.5}_{-0.0}$  &  0.19$^{+0.05}_{-0.04}$  
& 45$\pm4$ & 20$\pm2$ & I &     ... \\
Hen 3-1145 &  M1IVe &  3720  & 15.3$\pm0.1$  & 14.0$\pm0.1$ & -0.15$\pm0.15$  & 0.0$^{+0.1}_{-0.0}$ &   0.16$^{+0.05}_{-0.04}$  
& $<$7 & -2$\pm1$ &  I &    
[\ion{O}{i}] \\
\hline
\multicolumn{12}{c}{\it Giant stars with emission lines that may be Herbig Ae/Be stars}\\
\hline
HD 152291 &  B1IIIe &  24000  & 9.0$\pm0.1$   &  9.0$\pm0.1$   &  0.28$\pm0.15$  &  0.9$\pm0.5$  &  3.1$^{+1.3}_{-0.9}$  
& 320$\pm20$ & -20$\pm$10 &  II &     ... \\
Hen 3-416 &   A1IIIe &        9480  & 11.9$\pm0.2$   & 10.8$\pm0.1$   & 1.0$\pm0.2$  &  3.3$\pm0.7$  &  0.29$^{+0.12}_{-0.08}$  
& 30$\pm5$ & 33$\pm4$ &  III, IV &   [\ion{O}{i}] \\
WRAY 15-488 	&  F2IIIe 	 &    6870  & 10.3$\pm0.1$  & ~9.8$\pm0.1$	  & 0.11$\pm0.15$  &	 0.3$\pm0.5$  &	 0.36$^{+0.09}_{-0.07}$   
& 30$\pm5$ & -2$\pm2$ & IV &      ... \\
\hline
\multicolumn{12}{c}{\it Giant and Supergiant stars with emission lines}\\
\hline
HD 323154 			&  B4IIe		 &  15500  & 10.2$\pm0.1$  & ~9.3$\pm0.1$	&  1.10$\pm0.15$ &  3.4$\pm0.5$  &  1.1$^{+0.3}_{-0.2}$  
& 55$\pm5$ & 75$\pm$4 &  I &     ... \\
WRAY 15-1104 &  B1Iae	 &  20800  & 11.0$\pm0.1$  & 10.8$\pm0.1$ & 0.40$\pm0.15$  &  1.2$\pm0.5$  & 20$^{+7}_{-5}$  
& 50$\pm5$  & 60$\pm5$ & IV & [\ion{O}{i}] \\
Hen 3-1428 		&  B1Iae 	 &  20800  & 11.3$\pm0.1$  & 10.9$\pm0.1$ & 0.61$\pm0.15$  &  1.9$\pm0.5$  &  15$^{+6}_{-4}$  
& 38$\pm5$ & 30$\pm5$  & I &  [\ion{O}{i}] [\ion{N}{ii}] \\
MWC 314		&  B3Ibe 	 &  16200  & 11.1$\pm0.1$  & ~9.8$\pm0.1$	& 1.45$\pm0.15$  &  4.5$\pm0.5$  &  1.5$^{+0.4}_{-0.3}$  
& 40$\pm5$ & 65$\pm10$ & I & [\ion{O}{i}] [\ion{N}{ii}] \\
HD 320156		 &  B4Ibe	 &  14900  & 10.5$\pm0.1$  & ~9.9$\pm0.1$	&  0.72$\pm0.15$ &  2.3$\pm0.5$  &  4.2$^{+1.0}_{-0.8}$  
& 35$\pm5$  & 60$\pm3$ &  I & [\ion{O}{i}] [\ion{N}{ii}] \\
Hen 3-1347		&  B5Ibe &  13600  & 11.9$\pm0.2$  & 11.5$\pm0.2$	& 0.56$\pm0.28$  &  1.8$\pm0.9$  &  10$^{+6}_{-4}$  
& 30$\pm5$ & -15$\pm5$ & IV &    [\ion{O}{i}]  \\
AS 231				&  B9Iae &  10300  & 12.6$\pm0.6$  & 10.2$\pm0.1$	& 2.4$\pm0.6$   &  7.5$\pm1.9$  &  0.9$^{+1.3}_{-0.5}$ 
& 28$\pm5$ & -25$\pm5$& IV &    ... \\
\hline
\multicolumn{12}{c}{\it Stars with strong emission line spectrum}\\
\hline
WRAY 15-1651 &   B1-B5 &  ...  &  17.0$\pm0.2$  & 15.8$\pm0.2$  & ...  &  ...  &  ...  & ... &  ... & I &     [\ion{O}{i}] \\
MWC 930        &   B5-B9 &  ... &   14.6$\pm0.5$  & 12.8$\pm0.5$  & ...  &  ...  &  ...  & 70$\pm10$ &  10$\pm10$ & IV &    [\ion{O}{i}] \\
\hline
\multicolumn{12}{c}{\it Dwarf stars without H$\alpha$ in emission}\\
\hline
AS 321				&   A5V &   8200 & 12.0$\pm0.1$  & 11.3$\pm0.1$   & 0.60$\pm0.23$  &  1.9$\pm0.7$   &  0.31$^{+0.13}_{-0.09}$  
& 150$\pm10$ & -30$\pm5$ & abs. &  [\ion{O}{i}] \\
\hline
\multicolumn{12}{c}{\it Giant and Supergiant stars without H$\alpha$ in emission}\\
\hline
WRAY 15-522 &   G9III & 4820  & 14.3$\pm0.6$    & 13.4$\pm0.6$     & -0.03$\pm0.8$  &  0.0$^{+2.6}_{-0.0}$ &   3.3$^{+7.7}_{-2.3}$  
& $<7$ & -9$\pm2$ & abs. &  [\ion{O}{i}] \\
Th 35-41           &  M1III & 3720  & 14.5$\pm0.1$    & 13.1$\pm0.6$  & -0.17 $\pm0.6$ & 0.0$^{+1.5}_{-0.0}$ &   5.1$^{+5}_{-2.6}$  
& 13$\pm4$ &  15$\pm2$ & abs. &  ... \\
WRAY 15-566  &   M6III & 3240  & 16.0$\pm0.4$    & 14.2$\pm0.8$   & 0.31$\pm0.90$  &  1.0$\pm2.8$   &  4.6$^{+13}_{-3.5}$  
& $<$7 &  29$\pm2$ & abs. &  ... \\ 
WRAY 15-1650 &    M6III & 3240  & 13.5$\pm0.5$    & 11.9$\pm0.2$    & 0.13$\pm0.54$  & 0.4$\pm1.7$ &   2.1$^{+2.5}_{-1.1}$  
& $<$7 & -22$\pm2$ &  abs. &  ... \\
WRAY 15-1702 &    M6III & 3240  & 13.9$\pm0.2$    & 12.6$\pm0.2$    & -0.24$\pm0.29$  & 0.0$^{+0.1}_{-0.0}$ &   3.4$^{+0.4}_{-0.4}$  
& $<$7 &  -26$\pm2$ & abs. &  ... \\
WRAY 15-770 &   M7III & 3150  & 16.5$\pm1.0$    & 15.2$\pm1.0$    & -0.16$\pm1.4$  & 0$^{+3.9}_{-0}$ &   11$^{+62}_{-9.6}$         
& $<$7 & 23$\pm3$ & abs. &  ... \\
HD 305773 &  B1Ib &  20800  &  ~9.2$\pm0.1$    & ~9.1$\pm0.1$    &  0.27$\pm0.15$  &  0.8$\pm0.5$   &  6.6$^{+1.6}_{-1.3}$  
& 27$\pm$5 &  -2$\pm1$ & abs. & ...\\
\hline 
\end{tabular} 
\\[0.5cm]
\flushleft
Notes. 
Col.\,(3):  $T_{\rm eff}$ from Lang et et al. 1991 (adapted from Schmidt-Kaler 1982).\\  
Cols.\,(4)\,\&\,(5): photometry in the $B$ and $V$ bands taken from the NOMAD catalog,
except for Hen 2-80,  Th 17-35, and CD-38 4380 where they are from Vieira et al. (2003) and Hen 3-1121N where they are from the GSC 2.3.2.\\ 
Cols.\,(6)\,\&\,(7): Schmidt-Kaler (1982) intrinsic colors and absolute magnitudes of were used to calculate $E(B-V)$ and the distance.
The absolute magnitudes for the M6III and M7III stars are absent from the Schmidt-Kaler (1982) tables. 
We deduced them by linear extrapolation of the M$_V$ of the M3III and M4III spectral types. \\
Col.\,(6):$^\star$in the case of Hen 3-1121N the $R$=12.1$\pm 0.4$ band magnitude was used to derive $E(B-V)$ using
$E\,(B-V)=E\,(V-R)/0.78$.  \\
Col.\,(7): no correction for extinction was applied to sources with $E(B-V) < 0$.\\
Cols.\,(9) \& (10):$^*$when the spectra are dominated by emission lines and almost no absorption lines are observed
no constrain for $v ~{\rm sin}(i)$ is given and the  $v_r$ estimate is based on the averaged velocity shift of emission lines (principally \ion{He}{i}).\\
Col.\,(11): the type of the H$\alpha$ profile is given following the classification of Reipurth et al. (1996) and Vieira et al. (2003):
Type I profiles are symmetric without, or with only very shallow absorption features; Type II profiles are double-peaked
with the secondary peak having more than half of the strength of the primary; Type III profiles are double-peaked, 
with the secondary peak having less than half the strength of the primary; Type IV profiles have P Cygni line characteristics.\\
Col.\,(12):  F.L. means forbidden lines. 
[\ion{O}{i}] refers to the detection of [\ion{O}{i}] lines at 6301, 6365, or 8446~\AA; 
[\ion{S}{ii}] refers to the [\ion{S}{ii}] line at 6716 and 6731~\AA; [\ion{N}{ii}] refers to the lines at 6549 and 6585~\AA. 
\end{minipage}
\vfill
\end{table*}

\begin{table*}
\scriptsize	
\begin{minipage}[t]{\textwidth}
\caption{Equivalent widths of H$\alpha$,  \ion{He}{i},  \ion{Ca}{ii}, \ion{He}{i}, and forbidden emission lines observed in the spectra.}
\label{table:4}      
\centering                          
\begin{tabular}{l	c	c	c	c	c c c c c c c c c}
\hline
\hline
Star & H$\alpha$ & P(17) & \ion{He}{i} & \multicolumn{3}{c}{[\ion{O}{i}]} & \multicolumn{2}{c}{[\ion{S}{ii}]}  & \multicolumn{2}{c}{[\ion{N}{ii}]}
       & \multicolumn{2}{c}{\ion{Ca}{ii}$^\star$} \\
       & (6563\AA) & (8467\AA) & (6548\AA) & \multicolumn{3}{c}{(6301\AA,~6365\AA,~8446\AA)}  & \multicolumn{2}{c}{(6716\AA,~6731\AA)} 
       & \multicolumn{2}{c}{(6549\AA,~6585\AA)}
       & \multicolumn{2}{c}{(8498\AA,~8662\AA)}\\
\hline 
\multicolumn{13}{c}{\it Confirmed Herbig Ae/Be and T Tauri stars}\\
\hline
Hen 3-1121S      & EIA$^{+2.8}_{-0.1}$ 	& (0.1)	& +0.8 		& -0.02     & (0.01) & (0.05)	& (0.01) 	& (0.01) & +0.01 & (0.01) & (0.01) & (0.0) \\
Hen 3-1121N     & EIA$^{+2.0}_{-0.4}$ 	& (0.1)	& +0.8 		& -0.03     & (0.02) & (0.08)	& (0.01) 	& (0.01) & -0.04 	& -0.13 	& (0.1) & (0.0) \\
MWC 878          &  -54									& -2.0	&  -3.5 		&  -0.12    & (0.01) & -8.3		& (0.01) 	& (0.01) & -0.5 	& -1.6 		& -7.1   & -8.1 \\
WRAY 15-1435  & -20   								& -1.7	& +0.6	 	& -0.02     & (0.02) & -4.1		& -0.01   	& (0.01) & (0.01)	 & (0.02) & -0.3   & -0.5 \\
MWC 953          & -32   								& -3.5	& +0.3 		& -0.01     & (0.02) & -2.2		& (0.01) 	& (0.01) & +0.03 & +0.1 	& -0.4 	& -0.7 \\
HD 313571          & -34   								& -2.1	& -0.3		& -0.01     & (0.02) & -4.8		& (0.02) 	& (0.01) & +0.01 & (0.01) 	&  -1.3 	 & -0.1 \\
Hen 3-823          & -25   								& -1.4	& +0.6		& (0.01)   & (0.03) & -1.4		& (0.01) 	& (0.01) & (0.01) & (0.01) & (0.1)  & (0.1) \\
MWC 593          & -35  								& -1.3	& -0.2		& -0.01     & (0.03) & -3.7		& (0.01) 	& (0.01) & (0.01) & (0.02) &  -0.6	 & (0.1) \\
Hen 2-80            & -150								& -1.8	& +0.6		& -9.9       & -3.4     & -16			& -0.1  		& -0.2	  &  -0.06  & -0.2	 	& -0.2 	 & -0.2 \\
WRAY 15-1372  & -15									& (0.2)	& (0.02)	& (0.1)     & (0.1)  	& (0.6)		& (0.01)  	& (0.03) & (0.01) & (0.1) 	& (0.3)  & (0.6) \\
Th 17-35            & -90									& (0.6)	& +0.2		& -0.3       & -0.4    	& -5.3		& -0.1      & -0.1	  & -0.07 	 & -0.2 		& -0.1    & -0.2 \\
CD-38 4380       & EIA$^{+1.0}_{-3.0}$ 	& (0.6)	& (0.03)	& (0.1)     & (0.2)  	& +0.5		& (0.1)    & (0.02) & (0.01) & (0.3) 	& +0.7  & +1.4\\
HD 145718       & PC$^{+2.4}_{-0.2}$	& (0.6)	& (0.1)		& (0.2)     & (0.1)   	& +0.5		& (0.01) 	& (0.01) & (0.3) 	 & (0.0) 	& +0.6	 & +1.4\\
Hen 3-1145       & -30   								& (0.1)	& -0.5		& -0.5       & (0.3)  	& (0.02)	& +0.2    & (0.04) & (0.02) & (0.05)  & -0.5 	 & -0.5 \\
\hline
\multicolumn{12}{c}{\it Giant stars with emission lines that may be Herbig Ae/Be stars}\\
\hline
HD 152291      & -3.4								& -0.8	& +0.2				& (0.01) 	& (0.02)	& (0.05)	& (0.01) &  (0.01) 	& (0.01) & (0.01) 			& -0.5 & (0.1) \\
Hen 3-416        & PC$^{+0.5}_{-27}$	& (1.0)	&+0.2				& -0.4     	& -0.2     	& (0.0)		& (0.1)   &  (0.02) 	& (0.01) & (1) 				 	& -5.3 & -9 \\
WRAY 15-488  & PC$^{+7.2}_{-0.8}$& +1.0	& (0.06)			& (0.1)   	& (0.1)   	& +0.6		& +0.1   &  (0.1) 		& (0.01) & (0.3) 	 			& +1.2 & +2.0 \\
\hline
\multicolumn{13}{c}{\it Giants and Supergiants with emission lines}\\
\hline
HD 323154        & -26           					& -0.9	& -2.0				& (0.01) 	& (0.01)  & (0.02)	& (0.05) &  (0.01)  & (0.01) & -0.07 				& -0.8 & -0.9\\
WRAY 15-1104 & PC$^{+0.2}_{-10}$& +0.5	& PC$^{+0.4}_{-0.4}$	& (0.2)    & -0.02   	& -0.7		& (0.01) &  (0.01) 	& (0.01) 	& (0.3) 				
																																																																			& (0.5) & (0.5) \\
Hen 3-1428      & -18            					& PC$^{+0.25}_{-0.04}$ 
																							&  -1.4	   & -0.8   	& -0.3   	& -4.0		& (0.02) &   -0.05  	& -0.15		     & PC$^{+0.2}_{-0.9}$ 
																																																																			& (0.1) & (0.1)  \\
MWC 314        &  -118       					& -2.4	& -6.1     			& -0.01    & (0.01) 	& -4.7		& -0.06   &  (0.03) 	& -0.04 	& -0.17 			& -13 & -19 \\
HD 320156      & -7.8     							& -0.7	& 0.3			      & (0.02)  & (0.03) 	& PC$^{+0.2}_{-0.05}$		
																																												& (0.02) &  (0.02) 	& +0.1 		& PC$^{-0.15}_{+0.03}$ 
																																																																			& -8 		& -14 \\
Hen 3-1347      &PC$^{+0.8}_{-12}$  	& +0.1	& 0.2				 	& -0.3    	& -0.1	  	& -0.7		& (0.05) &  (0.04) 	& (0.01) 	& +0.2 				& +0.5 & EIA$^{+0.7}_{-0.2}$ \\

AS 231            & PC$^{+1.4}_{-14}$  & +0.7 	& 0.2				   & (0.01)  & (0.02)  & +0.8		& +0.01  & +0.03 	& (0.03) 	& (0.01)			& +1.3 & +2.0 \\
\hline
\multicolumn{13}{c}{\it Unclassified stars with strong emission line spectrum}\\
\hline
WRAY 15-1651 & -92  								& -1.2	& (0.6)			 & -0.90   & -0.1       & -7.4		& -0.1			 & -0.12 	& (0.06) & (0.02) 			& -1.5 & -1.7 \\ 
MWC 930         & PC$^{+0.1}_{-42}$& PC$^{+0.7}_{-1.0}$	& +0.2			 & -0.02   & (0.01)   & PC$^{+1.5}_{-1.5}$
																																														& PC$^{+0.2}_{-0.1}$& (0.04) & -0.04   & (0.01) 
																																																			 																& PC$^{+1.7}_{-5.1}$	 
																																																																			& PC$^{+2.0}_{-6.6}$ \\
\hline
\multicolumn{13}{c}{\it Dwarf stars without H$\alpha$ in emission}\\
\hline
AS 321             & +5.7 		& (0.1) 	& (0.03) 			& -0.02    & (0.01) 	& +0.6		& (0.01) 			&  (0.01) & (0.01) & (0.01) & +0.8 & +1.8\\
\hline
\multicolumn{13}{c}{\it Giant and Supergiant stars without H$\alpha$ in emission}\\
\hline
WRAY 15-522   & +1.2 		& +0.2		& (0.03) 	 & -0.2     & (0.1)    & (0.1)	  &  +0.1   &  (0.2) & (0.01)  & (0.1) & +1.3  & +2.1\\
Th 35-41           & +1.2			& +1.4		& -0.15		 & (0.1)   & (0.1)    & (0.2)	  & (0.5)    &  (0.4) & (0.01)  & (0.2) & +2.2 & +2.7 \\
WRAY 15-566   & +0.4			& (0.06)	& -0.15  	 & (0.3)   & (0.1)    & (0.1)	  & (0.6)    &  (0.1) & (0.1) 		& (0.2) & +0.6 & +0.6 \\
WRAY 15-1650 & +1.0			& +0.6		& (0.01) 	 & (0.2)   & (0.1)    & (0.2)	  & (0.5)    &  (0.2) & (0.1) 		& (0.3) & +1.3 & +2.1 \\
WRAY 15-1702 & +0.8			& +0.8		& (0.02) 	 & (0.1)   & (0.1)    & (0.3)	  & (0.5)    &  (0.5) & (0.1) 		& (0.3) & +1.1 & +2.0 \\
WRAY 15-770   & +0.5			& (0.05)	& (0.04) 	 & (0.2)   & (0.1)    & (0.1)	  & (1.0)    &  (1.0) & (0.1) 		& (0.1) & EIA$^{+0.4}_{-0.1}$ 
																																																									& PC$^{-0.3}_{+0.4}$  \\
HD 305773      & +2.0			& (0.07)	& +0.7	 	 & (0.05) & (0.01)  & (0.05) & (0.01)  &  (0.02) & (0.01)	& (0.05) & (0.1) & (0.1) \\
\hline
\end{tabular} 
\\[0.5cm]
\flushleft
Notes. In the cases where emission lines are observed inside a broad absorption line, we give the EW of the absorption and emission component and  
write EIA meaning ``Emission Inside Absorption".
In the case of P Cygni and inverse P Cygni profiles  are noted by PC and the equivalent width (EW) of the absorption
and emission component are given.  
For the stars displaying Paschen emission,
the \ion{Ca}{ii} lines at 8498 and 8662~\AA\ are blended with the Paschen(16) and Paschen(13) respectively.
For these sources the EW of \ion{Ca}{ii} line at 8498 and 8662~\AA, is the EW measured at the position of the \ion{Ca}{ii} line minus the 
EW of the Paschen(17) and Paschen(14) lines respectively. Upper limits are displayed in parenthesis.
\end{minipage}
\end{table*}

\subsection{Distance determination}
We estimated the distance of the sources based on spectroscopic parallaxes.
We  used the classical expression
$
{V}-M_{V}-A_{V}= 5 \,{\rm log}\,D  - 5
$,
where: $V$ is the observed $V$ magnitude, 
$M_{\rm V}$ is the intrinsic absolute magnitude corresponding to the spectral type and luminosity class 
found and $A_{\rm V}$ is the extinction.
$A_{\rm V}$ was calculated assuming the standard 
interstellar medium extinction law (Schultz \& Wiemer 1975)
$A_{\rm V}=3.14~E\,(B-V)$. 
For the stars with $B$ and $V$ magnitude measurements from different catalogs (i.e. different epochs)
but with $R$ band measurements from the same catalog, 
we employed $E\,(B-V)=E\,(V-R)/0.78$ (Schultz \& Wiemer 1975).
We employed the  intrinsic colors and absolute magnitudes (corresponding to each spectral type and luminosity class) 
of Schmidt-Kaler (1982)
and photometry from the literature (most notably the NOMAD catalog).
The error on $M_{V}$ and $(B-V)_0$ was set by the uncertainty of one spectral subclass in the spectral classification.
Some stars exhibited negative $E\,(B-V)$.
This is most likely due to photometry measurements taken at different epochs.
Herbig Ae/Be stars and post-main sequence supergiants show large variations in the optical, 
hence the need to take (quasi-)simultaneous photometry.
For these sources no correction for reddening was performed.
The resulting distances are given in Col. 8 of Table 3.

Given that the absolute magnitude for the luminosity class IV is not provided by the Schmidt-Kaler (1982) tables,
in the case of Hen 3-1145 (spectral type M1IVe),
we calculated the absolute magnitude ($M_{\rm V}$=7.9$\pm$0.6) from its $T_{\rm eff}$ (3720 K) and log$(g)$ (4.0 g/cm$^2$) (see Appendix 1).
We used the classical expressions $R=\sqrt{GM_{\star}/g}$, $M_{{\rm bol}}=M_{{\rm bol}\odot}-5{\rm~log}(R/R_{\odot})-10{\rm~log}(T_{\rm eff}/T_{\odot})$,
and $M_{\rm V}=M_{{\rm bol}}-BC$. We assumed a mass of 0.46$M_{\odot}$ for Hen 3-1145 (i.e. mass of a M1V star) and a bolometric correction (BC) of
-1.53 (i.e. median between the BC of the class V and class III for the M1 spectral type). 
The uncertainty in $M_{\rm V}$ being largely dominated by uncertainty on the value of log$(g)$. 
With an uncertainty in  log$(g)$ of 0.25 translating into an uncertainty of 0.6 mag in $M_{\rm V}$.

\subsection{Rotational velocity}
Employing our target - template interactive spectral comparison software, 
in a similar way as done with the spectral classification,
we compared  the continuum normalized FEROS spectrum with 
continuum normalized rotationally broadened theoretical spectra using
$T_{\rm eff}$ and log$(g)$ corresponding to the spectral type found for each of our stars.
The template spectrum was shifted wavelength in order to account for the
radial velocity of the target (see next section).
We used the theoretical spectral library BLUERED and in the case of early B-type stars the non-LTE B-star models TLUSTY.
We employed the $T_{\rm eff}$, log$(g)$ calibration tables of Lang et al. (1991).
The rotational broadening was calculated with the IDL routines
kindly provided by Bertone et al. in the BLUERED spectral library web-site.
For the analysis, once the BLUERED or TLUSTY  theoretical spectra were rotationally broadened,
they were degraded to the FEROS  resolution. 
For this, we used as well the Bertone et al. BLUERED IDL tools.

The projected rotational velocity uncertainty 
is the interval of  $v\, {\rm sin}(i)$ values that exhibited the smallest residuals and that best fitted 
the width and shape of the absorption lines of the FEROS spectrum. 
The uncertainty of $v\, {\rm sin}(i)$  was  typically 10\%,
the principal source of uncertainty being the value of $T_{\rm eff}$.
This being due to the fact that for a particular spectral type 
a range of $T_{\rm eff}$ exist that are consistent with it,
and for each  $T_{\rm eff}$  there is a range of $v\, {\rm sin}(i)$ that fit the spectra.
We present our results in Col. 9 of  Table 3.  

\subsection{Radial Velocity}
The FEROS pipeline effectuates the barycentric correction to the spectra (i.e. correction for the orbital motion of the Earth relative to the
solar system barycenter).
We derived the radial velocity of our targets with our interactive tool
by determining the velocity shift of our FEROS spectrum with 
respect to a rotationally broadened BLUERED or TLUSTY synthetic spectrum using
the $T_{\rm eff}$ and log$(g)$ of the spectral type and luminosity 
class found.
We proceeded by first continuum-normalizing our FEROS spectrum and the synthetic spectrum,
then interactively shifting the velocity of the FEROS spectrum until the center of the absorption lines of
target were at the same wavelength as the absorption lines of the synthetic spectrum
and the residuals between target and the synthetic spectrum were minimized. 
We used several regions in the spectra with strong absorption lines according to spectral type 
to obtain several estimations of the velocity shift.
The adopted velocity shift is the average of the velocity shifts found.
In the cases where the absorption lines are very weak, 
we employed Gaussian fits to the hydrogen or \ion{He}{i} emission lines to 
constrain the radial velocity.
We used the The Atomic Line List v2.4 website for the wavelength calibration\footnote{Hosted by the
Department of Physics and Astronomy,
University of Kentucky, and maintained by
Peter van Hoof
Royal Observatory of Belgium. http://www.pa.uky.edu/$\sim$peter/atomic/}. 
The typical error of the radial velocity is of a few km/s but can be as large as 10--20 km/s, 
depending on the spectral type and projected rotational velocity.
This is particularly relevant for B stars for which the velocity shift is mostly based on the \ion{He}{i} lines. 
Our results are summarized in Col. 10 of Table 3.

\subsection{Equivalent widths of selected emission lines.}
Employing our spectral-comparison software, 
we measured the equivalent widths of selected emission lines observed in the spectra:
H$\alpha$, Paschen (17),  \ion{He}{i} at 6548~\AA, \ion{Ca}{ii} at  8498~\AA\  and 8662~\AA,  
[\ion{O}{i}] at 6301, 6365 and 8446~\AA, [\ion{S}{ii}] at 6716 and 6731~\AA, and [\ion{N}{ii}] at 6549 and 6585~\AA.
We present the result our EW measurements in Table 4.

In the cases where emission lines are observed inside a broad absorption line 
or the lines display P Cygni or inverse P Cygni profiles,
we calculated individually the EW of the absorption and emission component.
To calculate the EW of P Cygni profiles, we first determined the continuum using a linear interpolation of the continuum
from nearby regions (right and left) outside the line. Then employing the standard EW formula we calculated the EW.
For the blueshifted absorption part, 
we integrated from the wavelength where the continuum and
left wing of the absorption line are similar (beginning of the absorption line), 
up to the wavelength where the right absorption wing crosses the continuum (end of the absorption line).
For the redshifted emission part, we integrated from the wavelength where the emission left wing 
crosses the continuum (beginning of the emission line) up to the wavelength where the continuum and
right wing of the emission line are similar (end of the emission line).

For the stars where Paschen (16) and Paschen (13) emission lines are blended with \ion{Ca}{ii} emission at 8498 and 8662~\AA,
we calculated the EW of \ion{Ca}{ii} lines by subtracting to the EW measured at the position of the \ion{Ca}{ii} line
the EW of the Paschen (17) and Paschen (14) lines respectively. 

\section{Results and Discussion}
The results of our FEROS spectroscopy campaign are summarized in Tables 3 and 4.
Table 3 describes the established spectral types, the estimated distance, the measured radial and projected rotational velocity, 
the type of the H$\alpha$ profile observed according to the classification of Reipurth et al. (1996) 
and a summary of important forbidden emission lines observed.
Table 4 displays the EW measurements of selected emission lines in the spectrum.

From our total sample of 34 candidates, 13 sources are confirmed as Herbig Ae/Be stars ($\sim$40\%).
These objects display H$\alpha$ in emission and have spectral types with luminosity class V.
They are mostly Herbig Be stars.
A large fraction of them (10 objects, $\sim$80\%) display forbidden line emission, in particular [\ion{O}{i}] lines (see Table 4).
\ion{Ca}{ii} emission lines at 8500~\AA\  are observed in $\sim$50\% of them (6 sources).
CD-38 4380 and HD 145718, the two new Herbig Ae/Be candidates from Vieira et al. (2003),
are confirmed as Herbig Ae/Be stars.
We observe (see Fig. 3) that, in general, for Herbig Ae/Be stars the H$\alpha$ line is centered close to the velocity of the star and that
its width is broader than the H$\alpha$ line observed in the supergiant stars of our sample.
Most notable exceptions are WRAY 15-1435 and HD 145718 where H$\alpha$ is observed shifted
with respect to the star's velocity and Hen 3-1121S\&N in which H$\alpha$ is narrow. 

From the subsample of 16 candidates positionally coincident with  nearby SFRs,
6 sources are confirmed as Herbig Ae/Be stars.
From these, two sources, CD-38 4380 and HD 145718, are at distances closer than 250 pc.
This provides further evidence for the association of CD-38 4380 with the Gum Nebula and of HD 145718 with Sco OB2.
The remaining 4 sources, Hen~2-80, Hen~3-823, Th~17-35, and WRAY~15-1372,
are at distances greater than 700 pc. They are not members of nearby SFRs. 

From the subsample of 18  ``isolated" candidates (i.e. candidates not known to be associated to a nearby SFRs),
we confirm 7 as Herbig Ae/Be stars: Hen 3-1121N, Hen 3-1121S, WRAY~15-1435, 
MWC~878, MWC~593, HD 313571, and MWC~953. 
All of them have distances greater than 700 pc. 

For the 11 confirmed Herbig Ae/Be stars with distances greater than 700 pc,  
we searched for Spitzer  8 $\mu$m images
and, using the SIMBAD database, for objects characteristic of SFRs
in their 20' vicinity (e.g., HII regions, molecular clouds, dark nebulae, maser emission, etc).
We wanted to check whether our distant Herbig Ae/Be stars  
are  truly ``isolated'' sources located in empty regions, 
or whether they may be associated with distant SFRs.
We found that seven sources 
(Hen~2-80, Hen~3--1121~N\&S, HD 313571, MWC~953,  WRAY~15-1435, and Th~17-35)
are  inside or  close ($<5'$) to regions with extended 
IR emission (see Fig. 6) and that in their 20' vicinity there are objects characteristic of SFRs.
As we do not have constrains on the distance of the extended IR emission, 
it is difficult to establish whether our sources are inside the emission region or if they are background or foreground sources.
However, as our stars are young, 
we find it is likely that they are inside these IR emission regions
and speculate that these regions may be distant SFRs. 
These regions are interesting for follow-up studies of their stellar content.
We remark that in the Spitzer images, three sources (WRAY 15-1435, HD 313571, Hen 2-80) 
are inside or have a nearby bright IR nebulosity.
This further confirms their young nature.
Two sources MWC 878 and WRAY 15-1372 do not have nearby extended IR-emission nor
nearby astronomical sources characteristic of SFRs. 
They may be a case of ``isolated" Herbig Ae/Be stars.
We note that Spitzer data were not available for MWC 593 and Hen 3-823.
In Appendix 2, we present the Spitzer 8 $\mu$m images and discuss each source individually.
 
Now, let us return to the discussion of the results of the complete sample.
One candidate, Hen 3-1145, is found to be an early M-type emission-line PMS star.
Li in absorption is observed towards this object confirming that it is a young classical T Tauri star.
The spectroscopic parallax distance of 160$^{+50}_{-40}$ pc strongly suggest its association to the Upper Centaurus Lupus 
SFR. 
Mer\'{i}n et al. (2008) observed this object with Spitzer and associated the source with the Lupus III dark cloud\footnote{
Mer\'{i}n et al. (2008) detected emission at 24, 70, and 160 $\mu$m in Hen 3-1145 with Spitzer. 
From the spectral energy distribution slope computed from the K-band to 24 $\mu$m, they classified the star as Class II.
The low IR excess led those authors to classify the source as an `anemic' disk:
it displays a 24 $\mu$m flux typical of Class III sources, but 70 $\mu$m excess comparable to a classical T Tauri star. 
Objects such as Hen 3-1145 appear to be extremely rare. They are the so-called cold disks
that are interpreted as optically thick disks with large inner holes of several to tens of AU (Calvet et al. 2005; Brown et al. 2007).}.

Three sources, HD 152291, Hen 3-416, and WRAY 15-488 have spectral types B, A, and F respectively,
luminosity class III and display H$\alpha$ and forbidden line emission (see Tables 3 and 4).
In these objects it is not entirely clear whether they are in the pre- or post-main sequence evolutionary phase.
Given that young early type stars are bright, they are able to dissipate their environment relatively early. 
It could be that HD 152291, Hen 3-416, and WRAY 15-488 are young stars still under contraction that have become optically visible. 
For that reason they display the signatures of low gravity, thus a luminosity class III.  
However, given that they could be post-main-sequence stars as well, we classified  them 
in the separate category ``Giant stars that may be Herbig Ae/Be stars".
The estimated distance can perhaps give some hints about the evolutionary stage of these sources.
In the case of HD~152291 the estimated distance of 3 kpc is {\it not} consistent with the association with the
SFR Upper Centaurus Crux (d$\sim$140 pc). Therefore, HD 152291 may be a background giant star.
Furthermore, the high value of $v~{\rm sin}(i)$ (320 $\pm$ 20 km/s) measured in HD 152291 makes  
it plausible that HD 152291 is a classical Be star, i.e. an evolved object spinning 
  near its break-up velocity, surrounded by a gaseous disk consisting of 
  matter lost by the central star.
In the case of Hen 3-416,  the estimated distance of 290$^{+110}_{-80}$ pc
does not provide strong evidence of the association to the SFR Scorpius OB 2-5 (d$\sim$140$\pm$20-30 pc, Preibisch and Mamajek 2008). 
But it might be that Hen 3-416 is a young star located at the very extreme end of the Scorpius SFR;
in this case Hen 3-416 is likely to be a Herbig Ae/Be star.  
In the case of  WRAY 15-488,  the estimated distance of 360$^{+90}_{-70}$ pc is more than 200 pc  
away than the SFR Scorpius OB 2-5. It is likely that the source is not associated to this SFR.
Present data do not permit us to conclude convincingly that WRAY 15-488 is a young star.

Seven of our 34 sources (20\%) are post- main sequence supergiant stars {\it with} H$\alpha$ in emission.
These seven objects are not members of the Herbig Ae/Be stellar group.
High-resolution spectra were able to reveal the spectral diagnostics needed to establish their evolved state. 
All of them display H$\alpha$ in emission, and  forbidden line emission was observed in five cases (WRAY~15-1104, Hen~3-1428, MWC~314,
HD 320156, and Hen~3-1347).
These objects are examples of post-main sequence B[e] stars.
Almost all of them display P-Cygni Balmer emission line profiles or blueshifted (50-150 km/s) H$\alpha$ emission.
This suggests the presence of strong winds.
We observe (see Fig. 3) that, in general, the width of their H$\alpha$ line is narrower than the H$\alpha$ emission line observed 
in the Herbig Ae/Be stars of our sample.
Although not in the direction of our goal of finding young stars, these objects are very interesting because 
they are examples of B[e] supergiants with IR excess.

Two sources (WRAY 15-1651, MWC 930) display rich emission line spectra. 
Their spectra  exhibit strong veiling,  likely due to accretion. 
For these sources we were unable to derive their luminosity class because 
of the nearly complete absence of photospheric absorption lines in their optical spectra. 
We assume they are {\it bona-fide} young stars, but we were not able to constrain their distances.

One source (AS 321) is an A-type main sequence star without emission lines. 
This object is located at a distance $\sim$300 pc and it is not associated with a known nearby SFR.
It might be a field A star. The lack of emission can be understood trough two scenarios.
In the first scenario the star is a main-sequence star, and the measured IRAS excess is due to a debris disk surrounding the
star (i.e. a vega-like star). 
In another scenario, AS 321 is a young Herbig Ae star, 
but due to the intrinsic variability of accretion,
the H$\alpha$ line is absent because  
we observed AS 321 during a period of low accretion activity.
We should note that AS 321 does exhibit [\ion{O}{i}] forbidden line emission at 6300~\AA\ and 
this could be due to the presence of either a 
disk or an outflow\footnote{A recently acquired high-resolution optical spectrum of AS 321 
with UVES at the VLT (van den Ancker et al., in prep.) confirms the absence 
of H$\alpha$ emission in the optical spectrum of AS 321.}.

Finally, seven of our 34 sources (20\%) are giant  (6) and supergiant (1) stars {\it without} 
H$\alpha$ in emission (WRAY~15-522, Th~35-41, WRAY~15-566, WRAY~15-1650, WRAY~15-1702, WRAY~15-770, HD 305773).
They are rejected as Herbig Ae/Be stars because they are evolved objects.
Most of them have a late spectral type (i.e. M). 
Their distances indicate that they are field-background stars.
Since most of these stars originate from surveys of emission-line stars, 
the lack of H$\alpha$ in emission was not expected.  
As in the case of AS~321, there is always the possibility of spectral variability of the H$\alpha$ line.
However, in some of these cases, the target may also simply have been misidentified, as the accuracy 
of coordinates in the catalogs from which these targets were selected is low.
We selected the brightest K-band source in the vicinity of the coordinates given in the Th\'e et al. (1994) catalog,
but it may be that the emission line-star is a fainter IR source in the field.  

\section{Summary and Conclusions}
We obtained high-resolution optical spectroscopy 
of 34 candidates to Herbig Ae/Be stars with unknown or poorly constrained
spectral types from the Th\'e et al. (1994) catalog and two candidates from Vieira et al. (2003).
We observed 16 candidates positionally coincident with nearby (d$<$250 pc) SFRs
and 18 relatively bright ($V < 14$) ``isolated" candidates.
All our candidates have reported IR-excess from IRAS.
Our aim was to determine whether the candidates are Herbig Ae/Be stars or
background giants, and in the specific case of the candidates positionally coincident with SFRs,
we wanted to further find out whether they are members of the SFR.
We determined the spectral types of the sources by careful comparison with spectral templates,
we measured their radial and projected rotational velocities,
finally, we constrained their distances employing spectroscopic
parallaxes based on the intrinsic colors of the established spectral type and luminosity 
class and photometry from the literature.

From the 34 Herbig Ae/Be candidates studied, 
26 objects exhibit H$\alpha$ in emission ($\sim$80\%, see Figure 3 and Table 3).
From these 26 objects, 
14 are dwarfs and subgiants (luminosity classes V and IV),
10 are giants and super giants (luminosity classes III, II, and I),
and 2 are unclassified extreme emission line objects.
From the 8 objects {\it without} H$\alpha$ emission,
7 are giants and one (AS 321) is 
a main-sequence A-type star.

Among the 14 emission line dwarfs and subgiants, 13 objects are confirmed Herbig Ae/Be stars and one is a CTTS.
In addition to these 13 confirmed Herbig Ae/Be stars, 5 additional 
sources might be Herbig Ae/Be stars: 3 emission-line early type luminosity class III giants, 
and 2 extreme emission line objects. 
However, our data did not allowed us to firmly establish whether these 5 sources are truly Herbig Ae/Be stars.

Two confirmed Herbig Ae/Be stars (CD-38 4380, HD~145718) and the CTTS (Hen 3-1145)
are at distances closer than 250 pc. These sources are likely members of nearby SFRs.
One emission line giant star (Hen 3-416) is at closer than 300 pc.
If this source is a young star, it may be associated with Sco OB 2-5.
These 4 sources are likely to be nearby young stars and are interesting for  follow-up observations at high-angular resolution.
The rest of our confirmed Herbig Ae/Be stars (11 sources) are at distances greater than 700 pc.
From this subsample, 7 stars (Hen~2-80, Hen~3--1121~N\&S, HD 313571, MWC~953,  WRAY~15-1435, and Th~17-35) 
are inside or close (separation $<5'$) to regions with extended infrared (IR) continuum emission at 8$\mu$m
and have astronomical sources characteristic of SFRs in their 20' vicinity.
These 7 sources are likely to be members of distant SFRs.
Such regions are attractive for future studies of their stellar content.
Two confirmed Herbig Ae/Be stars at $d>700$ pc, MWC 878 and WRAY 15-1372,
may be truly ``isolated" sources.

From our 34 Herbig Ae/Be candidates we found that $\sim$50\% (15 of 34) turned out to be background giant stars and not young stars.
They show us that high-resolution optical spectroscopy is an important tool for distinguishing young stars (in particular Herbig Be stars) from 
post-main sequence stars in samples taken from catalogs based on low-resolution spectroscopy.
A systematic study of large samples of candidates to Herbig Ae/Be stars employing
high-spectral resolution spectroscopy is fundamental for firmly establishing their genuinly young nature.
      
\begin{acknowledgements}
M.A and A.C acknowledge support from a Swiss National Science Foundation grant (PP002--110504).
A.C. would like to thank C.~Baldovin-Saavedra and F. Fontani for comments to the manuscript,
J. Hernandez for discussions about the spectral classification of Herbig Ae/Be stars,
M. Chavez for providing a DVD with the UVBLUE/BLUERED spectral library,
and A. Mueller for kindly making a set of high-resolution Kurucz synthetic spectra of low-mass stars available. 
We acknowledge the ESO and MPIA staff for carrying out the observations.  
This research made use of Aladin, the SIMBAD database and the VizieR service
operated at the CDS, Strasbourg, France.
This work is based in part on observations made with the Spitzer Space Telescope, which is operated by the Jet Propulsion Laboratory, California Institute of Technology under a contract with NASA.
\end{acknowledgements}

\section*{Appendix 1: Comments on the spectral classification of individual sources.}  
\subsection*{AS 231}
In AS 231 we observe H$\gamma$, H$\beta$, H$\alpha$,  \ion{He}{i}, \ion{Ca}{i} (3933~\AA), 
\ion{Fe}{ii}  (4233, 4352~\AA), and \ion{Ti}{I} (4550, 4583~\AA) P-Cygni profiles.
Given that \ion{He}{ii} absorption lines are absent in the spectrum, 
the presence of a few \ion{He}{i} lines in absorption indicates that AS 231 is a B star.
As the \ion{Mg}{ii} line at 4481~\AA\  is much stronger than the \ion{He}{i} line at 4471~\AA
and the \ion{He}{i} line at 4009~\AA\  is absent from the spectrum,
AS 231 should have a spectral type later than B8.
Since \ion{Si}{ii} lines are observed at 3854, 3856, 3863, 4128, and 4131~\AA,
AS 231 has luminosity class I.
The absence of the \ion{He}{i} line at 4144~\AA, 
the lack of strong \ion{C}{ii} line at 4267~\AA\ 
and the presence of \ion{Fe}{ii} in absorption
at 4173, 4179, and 4417~\AA, rule out the spectral type B8, and indicate that AS 231 
should have a spectral type B9.
The presence of a relatively strong \ion{Fe}{ii} lines at 4179, 4233, and 4352~\AA\  further indicates that 
the luminosity class is Ia.
We conclude that AS 231 has a B9Iae spectral type.

\subsection*{AS 321}
AS 321 does not show hydrogen emission lines in our spectrum.
Hydrogen is observed in absorption and the profiles are broad.
A strong and broad \ion{Ca}{ii} K line in absorption is present.
These characteristics indicate that AS 321 is an A-type star.
The width and the strength of the \ion{Ca}{ii} K line and the hydrogen lines indicate that 
AS 321 is a mid-A star.
The width of the hydrogen lines, the width and shape of the Fe absorption lines,
the broad \ion{Ca}{i} line at 4227~\AA\ and the absence of lines such as the \ion{Ti}{ii} lines at 3901 and 3913~\AA,
rule out the luminosity classes I and II.
The profile's shape of the Fe lines can be reasonably matched with the spectrum of A7III and A8III stars
once rotational broadening is taken into account.
However, since the hydrogen lines profiles obtained are narrower than those of our FEROS
spectrum, we deduce that AS 321 should be of luminosity class V.
The spectral template that best fits the observations of AS 321 is the one of an A5V star.
Therefore, we conclude that the spectral type of AS 321 is A5V.

\subsection*{CD-38 4380}
CD-38 4380 exhibits H$\alpha$ in emission.
Its spectrum does not exhibit \ion{He}{i} or \ion{He}{ii} absorption lines.
Several weak absorption lines are observed in between the Balmer lines.
This combined with the strength of the \ion{Ca}{ii} K line at 3933~\AA\  indicates  that  CD-38 4380 must be later than A9.
The lack of a strong G-band feature at 4300~\AA\  shows that  CD-38~4380 is earlier 
than G-type.
Therefore, CD-38 4380 is most likely a F-type star. 
The width of the Balmer lines at 3889 and 3970~\AA\  further indicates that CD-38 4380 should have a spectral type later than F3.
The strength of the \ion{Ca}{i} line at 4227~\AA, the \ion{Fe}{i} lines at 4005, 4144~\AA,
and the \ion{Sr}{ii} line at 4078~\AA\  line indicates that CD-38 4380 should be of spectral type earlier than F6.
To determine the luminosity class we analyzed the region around the \ion{Ca}{i} line at 4227~\AA.
The strengths the \ion{Sr}{ii} line at 4216~\AA, the \ion{Ca}{i} line at  4227~\AA, the \ion{Fe}{ii} line at 4247~\AA,  
and other absorption lines observed in the region are too weak to be compatible with the luminosity classes I and II.
Templates of F2 to F4 class III giants  are compatible with the general shape of the observed spectrum.
However, the best match is provided by a rotationally broadened ($v$ sin$(i)$~45 km/s) F4V template.
Thus, we conclude that CD-38 4380 has a F4Ve spectral type.
Previous studies based on low resolution spectra suggested spectral types  F3V  (Vieira et al. 2003, PDS 277) and
 F2Ie (Suar\'ez et al. 2006).
Our data rule out the F2 supergiant spectrum. 
F3V is compatible with the observed data, but our high resolution
spectrum is better described by a F4Ve spectrum.

\subsection*{HD 145718}
HD 145718 displays an inverse P Cygni H$\alpha$ profile. 
Other Balmer lines are observed in absorption, but their profiles are not symmetric.
The blue part displays a shoulder consistent with filling with an emission component.
The spectrum lacks of \ion{He}{i} lines and a broad \ion{Ca}{ii} K line at 3933~\AA\  is observed. 
The strength the \ion{Ca}{ii} K line shows that HD 145718 is an A-type star. 
Its width indicates that HD 145718 has a spectral type later than A3 but earlier than A7.
The strength of the \ion{Mg}{ii} line at 4481~\AA\  constrains the spectral type to be earlier than A6.
The strength of the \ion{Ca}{i} line at 4227~\AA\  constrains the spectral type to be later than A4.
Comparison with the spectral templates shows that the spectral type that best match the
observed spectrum is A5.
The weak strength of the \ion{Fe}{ii} - \ion{Ti}{ii} lines at 4173~\AA, the \ion{Y}{ii} - \ion{Fe}{ii} lines at 4179~\AA, 
the \ion{Ti}{ii} lines at 3901 and 3913~\AA\ 
rules out the luminosity classes I and II.
Analyzing the region between the H$\gamma$ line and the \ion{Ti}{ii} line at 4583~\AA, 
we found that the width of the weak absorption features visible in the spectrum are much
better matched by a dwarf star (luminosity class V) than with a rotationally broadened giant (luminosity class III).
Therefore we conclude that HD 145718 is a dwarf star and has a spectral type A5Ve.
HD 145718 was studied by Guimaraes et al. (2006) who found a $T_{\rm eff}=7500\pm200$K,
corresponding to a somewhat later spectral type than the A5Ve found by us.
However, we note, as previously mentioned, that the strength of the \ion{Mg}{ii} line at 4481~\AA\  indicates that the spectral type is earlier than A6.
Comparison of the 100 \AA\ region around the \ion{Mg}{ii} line at 4481~\AA\  with BLUERED synthetic spectra indicates that the $T_{\rm eff}$
that best matches our HD 145718 spectrum is in the 8000-8500 K range (for log~$g$ ranging from 3.0 to 4.0). 
This further indicates that the spectral type of HD 145718 is A5Ve. 

\subsection*{HD 152291 = MCW 1264}
HD 152291 displays H$\alpha$ in emission.
Its spectrum exhibits \ion{He}{i} in absorption. 
Given that no \ion{He}{ii} absorption lines are observed, HD 152291 should have a spectral type later than B0.5.
The presence of the \ion{He}{i} line at 4121~\AA\  and a weak \ion{Mg}{ii} line at 4481~\AA\  
indicates that HD 152291 is earlier than B5.
The presence of a strong \ion{O}{ii} - \ion{C}{iii} blend at 4650~\AA\   shows that HD 152291 
is earlier than B2.
The presence of the \ion{O}{ii} lines at 4070, 4300 and 4415~\AA\  rules out the
luminosity class V.
The lack of the \ion{N}{ii} lines at 3995 and 4631~\AA,  and the presence of strong \ion{Si}{iii}  triplet at 4553~\AA\  
rule out the luminosity class I - II.
Therefore, we conclude that HD 152291 has a spectral type B1IIIe.

\subsection*{HD 305773}
HD 305773 does not show hydrogen emission lines.
\ion{He}{i} absorption lines at 4026, 4144, 4471, 5016, 5876, 6678, and 7066~\AA\  are observed.
Since \ion{He}{ii} lines are absent, HD 305773 should have a spectral type B.
The presence of the \ion{He}{i} line at 4009, 4121, and 4388~\AA\  indicates that the spectral type is 
earlier than B5.
The detection of the \ion{N}{ii} line at 3995~\AA, the \ion{O}{ii} lines at 4070 and 4976~\AA, 
the \ion{Si}{iii}  lines at 4553, 4568, and 4575~\AA\  indicates that the luminosity class of 
HD 305773 is I.
The lack of the \ion{Si}{ii} lines at 4128 and 4131~\AA\  shows that HD 305773 is earlier than B3.
The presence of the \ion{Si}{iv} lines at 4089 and 4016 ~\AA\  indicates that HD 305773 should be
of spectral types B0-B1.
The presence of a rich spectra of \ion{O}{ii} lines and the strength of the \ion{Mg}{ii} line at 4481~\AA\  
constrain the spectral range to be B1 or later.
Therefore, HD 305773 should have a spectral type B1. 
The detection of the \ion{C}{ii} line at 4267~\AA\  indicates that the luminosity class is Ib.
In summary, HD 305773 has a B1Ib spectral type.

\subsection*{HD 313571 = MWC 595}
The spectrum of HD 313571 shows H$\alpha$, H$\beta$, H$\gamma$, and H$\delta$ in emission.
Double peaked hydrogen Paschen lines are  observed in emission at 8000~\AA.
The \ion{Ca}{ii} triplet at 8500~\AA\  is observed in emission superposed to the hydrogen Paschen lines.
\ion{He}{i} is observed in absorption (e.g., at 4026, 4144, 4121, 4386, 4471, and 4713~\AA) and in emission 
(e.g., 4922, 5016, 5876, 6678, and 7066~\AA).
Since \ion{He}{ii} lines are not present in the spectrum, HD 313571 is a B-type star.
The lack of the \ion{He}{ii} line at 4686~\AA\  shows that HD 313571 has a spectral type later than B0.5.
The \ion{Mg}{ii} line at 4481~\AA\  is present and it is weaker than the \ion{He}{i} line at 4471~\AA,
thus HD 313571 should have a spectral type earlier than B6.
The strength of the \ion{He}{i} line at 4009~\AA\  indicates that HD 313571 should have a spectral type B3 or earlier.
The absence of the \ion{N}{ii} line at 3995~\AA, the \ion{Si}{ii} lines at 4128 and 4131~\AA, the \ion{Si}{iii}  lines at 4553, 4568, and 4575~\AA,
the \ion{Si}{iv} line at 4089 and 4116~\AA, and \ion{O}{ii} lines in the regions at 4000, 4300, and 4400~\AA,
indicates that HD 313571 is not a supergiant star (luminosity classes I and II).
The lack of \ion{O}{ii} absorption at 4415 and 4417~\AA\  rules out the spectral type B1 III.
The strength of the \ion{Mg}{ii} line at 4481~\AA\  suggests a spectral type later than B1.
The lack of \ion{Si}{iii}  lines at 4553, 4568, and 4575~\AA\ rules out the spectral type B2III.
The absence of the \ion{Si}{ii} lines at 4128 and 4131~\AA\  indicates a spectral type later than B2V.
The strength of the \ion{He}{i} line at 4009~\AA\  with respect to the \ion{He}{i} line at 4026~\AA\ 
and the presence of a the \ion{C}{ii} line at 4267~\AA\ rule out the spectral types B4III-V.
Comparison with spectral templates shows that the spectrum that better matches the relatively intensity of the \ion{He}{i} lines is
the spectral type B3V.
We conclude that HD 313571 has a B3Ve spectral type.
 
\subsection*{HD 320156 = Hen 3-1444}
HD 320156 displays H$\alpha$ and H$\beta$ in emission.
Inverse P Cygni  \ion{He}{ii} profiles are observed. 
The \ion{He}{ii} line at  4686~\AA\  is absent.
A rich spectrum of \ion{He}{i} lines is observed.
The \ion{He}{i} lines are narrow, typically of {\it FWHM} $\sim$ 45 km/s,
some of them display inverse P Cygni profiles, and in some cases double peaked profiles.
The \ion{Mg}{ii} line at 4481~\AA\  is present and has a strength similar to the
\ion{He}{i} line at 4471~\AA.
This ensemble of characteristics suggests that HD 320156 is an early B-type star.
The absence of the \ion{He}{ii} line at  4686~\AA\   indicates that HD 320156 
has a spectral type B1 or later.
An \ion{He}{i} line at 4009~\AA\  of similar strength than the \ion{He}{i} line at 4026~\AA\ 
narrows down the spectral types to B2 to B4.
The spectrum displays \ion{C}{ii} lines at 3919 and 3921~\AA, \ion{N}{ii} at 3995~\AA,
\ion{Si}{ii} at 4128 and 4131~\AA, and \ion{Si}{iii}  at 4553, 4568, and 4575~\AA.
This, combined with the narrow H lines observed in absorption, rules out the luminosity classes II, III, and V,
and shows that HD 320156 is a supergiant star of luminosity class I.
The presence of the mentioned \ion{Si}{ii} and \ion{Si}{iii}  lines, and the lack of strong \ion{O}{ii} lines at 4070, 4076, 4346, and 4649~\AA\ 
indicate that  its spectral type should be B3 to B4.
The similar strength of the \ion{He}{i} line at 4121~\AA,  and the \ion{Si}{ii} lines at 4128 and 4131~\AA\ 
indicates a luminosity class Ib and suggests a spectral type B4.
We conclude that HD 320156 has a spectral type B4Ibe.
Finally, we note that Sch\"onberner \& Drilling (1984) suggested that HD 320156 (= LSS 4300) is a close 
binary system consisting of a helium supergiant of $\sim$ 1~M$_\odot$ and a less luminous secondary.  
Our FEROS spectrum is compatible with this hypothesis.

\subsection*{HD 323154 = MWC 877}
HD 323154 displays a flat spectrum characterized by the presence of \ion{He}{i} in absorption
at 3820, 4009,  4121, and 4713~\AA.
\ion{He}{i} is observed in emission starting with the line at  4922~\AA.
No \ion{He}{ii} lines are present in the spectrum.
Hydrogen lines are observed in emission starting with the H$\delta$ line at 4102~\AA.
The H$\alpha$ line exhibits a double peaked profile.
Broad hydrogen Paschen lines are observed in emission starting at 8360~\AA.
The lack of \ion{He}{ii} lines and the presence of \ion{He}{i} lines indicate that HD 323154 is a B star.
The \ion{Mg}{ii} line at 4481~\AA\  has an absorption depth slightly weaker than the \ion{He}{i} line at 4471~\AA.
This indicates that HD 323154 has a spectral type earlier than B6 (given that the \ion{He}{i} profile is contaminated by emission,
we can only use the \ion{Mg}{ii}/\ion{He}{i} ratio to set an upper spectral type limit).
The presence of \ion{N}{ii} absorption at 3995~\AA\  rules out the luminosity class V and IV for HD 323154,
and indicates that HD 323154 has a spectral type earlier than B5.
The weak \ion{C}{ii} line at 4267~\AA\  and  the lack of \ion{O}{ii} lines at 4070, 4076, 4346, 4349, 4415, and 4417~\AA\  
rule out the spectral types  B1I-III and B2I-III.
The strength of the \ion{C}{ii} line is weaker than observed in B3I-III stars, thus  HD 323154 should have a spectral type B4.
A strength of the \ion{Mg}{ii} line similar to that of the \ion{He}{i} 4471 line and the strength of  \ion{Si}{ii} lines at 4128 and 4131~\AA\   rule out the
luminosity class I. The presence of \ion{Si}{iii}  absorption at 4553~\AA\  and the \ion{N}{ii} line at 4631 are inconsistent with the
luminosity class III. Therefore, the luminosity class of HD 323154 should be II.
We conclude that HD 323154 has a spectral type B4IIe.
 
\subsection*{Hen 2-80}
The spectrum of Hen 2-80 is flat and it exhibits only a few absorption lines. 
The presence of weak \ion{He}{i} lines at 4026, 4144, and 4471~\AA\ 
suggests that Hen 2-80 has a spectral type earlier than B8.
Since the \ion{Mg}{ii} line at 4481~\AA\  is just slightly weaker than the \ion{He}{i} line at 4471~\AA, 
Hen 2-80 should have a spectral type B5 or B6.
The absence of the He 4009~\AA\  and the presence of the \ion{He}{i} at 4026~\AA\  
suggest that Hen 2-80 has a spectral type B6 or later.
The lack of the \ion{Si}{ii} triplet at 3855~\AA\  rules out the luminosity classes I and II.
Given that the \ion{Mg}{ii} line width is much narrower than the \ion{He}{i} line at  4471~\AA,
Hen 2-80 is not of luminosity class III.
We conclude that Hen 2-80 has a spectral type B6Ve.

\subsection*{Hen 3-416}
Hen 3-416 exhibits P-Cyni profiles in the Balmer lines.
Strong H$\alpha$ in emission is observed.
\ion{Ca}{ii} emission at 8498, 8542, and 8662~\AA\ is observed as well. 
\ion{He}{i} lines in absorption are not present in the spectrum.
However, the \ion{He}{i} lines at 4922 and 5016~\AA\  display P-Cygni profiles.
The lack of \ion{He}{i} absorption points to a star of spectral type A or later.
The lack of the \ion{Ca}{i} line at 4227~\AA\  indicates that Hen 3-1416 should be of spectral type
earlier than A3. 
A narrow and relatively strong \ion{Ca}{ii} K line at 3933~\AA\  and the strength of the \ion{Mg}{ii} line at 4481~\AA\ 
are consistent with a spectral type earlier than A2.
The very narrow Balmer lines indicate that Hen 3-416 is not a dwarf (i.e. not of luminosity class V).
The lack of multiple absorption lines in the region of around the \ion{Mg}{ii} line at 4481~\AA\  and the \ion{Ca}{i} line at 4227~\AA\ 
rules out the luminosity classes I and II.
The strength of the \ion{Fe}{ii} line at 4233~\AA\  shows that the spectrum is later than A0.
The lack of \ion{Ca}{i} absorption at 4227~\AA\  rules out the spectral type A2III.
This ensemble of diagnostics suggests that Hen 3-416 has a spectral type A1III.
We conclude that Hen 3-416 has an A1IIIe spectral type.

\subsection*{Hen 3-823 = CD-59 4412}
Hen 3-823 displays broad H$\alpha$ and \ion{Ca}{ii} in emission.
Its spectrum is flat and exhibits several lines of \ion{He}{i}.
Since \ion{He}{ii} absorption is not observed, in particular the line at 4686~\AA,
Hen 3-823 should have a spectral type B later than B0.5.
The presence of the \ion{He}{i} lines at 4009, 4121, 4026, 4388, and 4471~\AA\ 
and a \ion{Mg}{ii} line at 4481~\AA\  weaker than the \ion{He}{i} line at 4471~\AA\ 
indicate that Hen 3-823 should have a spectral type earlier than B5.
The strength of the \ion{Mg}{ii} and the \ion{He}{i} line at 4009~\AA\  suggests that Hen 3-823
should have a spectral type B2 or later.
The strength of the \ion{He}{i} lines at 4121~\AA\  and 4388~\AA\  shows that Hen 3-823
should have a spectral earlier than B4.
The lack of the \ion{Si}{iii}  lines at 4553, 4568, and 4575~\AA,  and \ion{O}{ii} lines at 4070, 4349, and 4415~\AA\ 
rules out the luminosity classes I and II.
Broad \ion{He}{i} absorption profiles and the absence of  \ion{He}{i} at 3927~\AA\
rule out the luminosity class III and indicate that Hen 3-823 is a dwarf.
The strength of the \ion{He}{i} lines is best matched by a B3 star.
We conclude Hen 3-823 has a spectral type B3Ve.

\subsection*{Hen 3-1121N}
Hen 3-1121N exhibits a weak (EW$=-0.3$~\AA) narrow ({\it FWHM} = 35 km/s) H$\alpha$ line in emission  
observed inside a broad H$\alpha$ line in absorption ({\it FWHM} = 460 km/s).
The spectrum is rich in \ion{He}{i} absorption lines.
Since \ion{He}{ii} at 4686~\AA, \ion{C}{iii} at 4647 and 4651~\AA, \ion{C}{ii} at 4267~\AA,  
and \ion{Si}{iv} at 4089 lines
are observed in absorption, Hen 3-1121N should have a spectral type earlier than B1.
The lack of the \ion{He}{ii} line at 4542~\AA\  rules out the late O spectral types and
indicates that Hen 3-1121N has an early B spectral type. 
Presence of the \ion{Si}{iv} line at 4089~\AA, the lack of the \ion{Si}{iv} line at 4116~\AA\ 
and the lack of strong \ion{Si}{iii}  lines at 4553, 4568, and 4575~\AA\  suggest that 
Hen 3-1121N is of spectral type B0.5.
The lack of \ion{O}{ii} absorption lines at 4317, 4320, 4346, 4349, 4367, 4415, and 4417~\AA\ 
rules out the luminosity classes I, II, and III. 
We conclude that Hen  3-1121N has a spectral type B0.5Ve.

\subsection*{Hen 3-1121S}
Hen 3-1121S displays a broad H$\alpha$ line, 
a very weak emission component (EW$=-0.05$~\AA) is observed inside the absorption line.
The spectrum exhibits a rich spectra of \ion{He}{i} lines. \ion{He}{ii} is observed in absorption at 
4686~\AA\  and no other \ion{He}{ii} lines are observed.
\ion{C}{iii} absorption is observed at 4647 and 4651~\AA.
These spectral characteristics indicate that Hen 3-1121S has a spectral type B earlier 
than B1.
The \ion{Si}{iv} line at 4089~\AA\  is stronger than the \ion{Si}{iii}  line at 4552~\AA\, indicating that the spectral type is B0. 
The presence of the \ion{He}{ii} line at 4686~\AA\  combined with
the width and shape of the hydrogen Balmer lines and the absence of \ion{O}{ii} lines at 4317, 4320, 4346, 4349, and 4367~\AA,
and the \ion{N}{ii} line at 3995~\AA\  indicate that the star {\it is not} of luminosity class I or II.
The strength of the \ion{He}{i} 4009 relative to the strength of the \ion{He}{i} 4026~\AA\  line 
indicates that Hen 3-1121S is of luminosity class V.
We conclude that Hen 3-1121S has a spectral type B0Ve.

\subsection*{Hen 3-1145}
 Hen 3-1145 displays a spectrum rich in emission lines.
 We observe in emission hydrogen Balmer lines
 (H$\alpha$, H$\beta$, H$\gamma$, H$\delta$, H$\upsilon$, H$\xi$),
 \ion{He}{i} at 4026, 4471, and 5876~\AA, \ion{He}{ii} at 4686~\AA, 
 \ion{Ca}{ii} at 3933~\AA\  and the \ion{Ca}{ii} triplet at 8500~\AA.
Li is observed in absorption at 6708~\AA\  (EW = 0.5~\AA).
The spectrum is also rich in narrow absorption lines, thus Hen 3-1145 should have a late spectral type.
The strength and width of the \ion{Ca}{i} line at 4227~\AA\ shows that Hen 3-1145 has a spectral type K or later 
(note that the \ion{Ca}{i} line is likely to be contaminated by emission; thus, the real \ion{Ca}{i} strength is intrinsically larger than
observed).
The presence of weak molecular bands at 7000~\AA\  indicates that Hen 3-1145 has a spectral type later than K7
but earlier than M2 (otherwise the molecular bands will be much stronger).
The strength of the \ion{Na}{i} D lines at  5890~\AA\  and the molecular bands at 7000~\AA\  
indicate that Hen 3-1145 has a spectral type M0 to M1.
The smallest residuals to the molecular bands at 7000~\AA\   are given by the spectral type M1.
The absence of the absorption doublet at 5860~\AA\  further indicates that the spectral type should be M1.
The broad width of \ion{Na}{i} D lines rules out the luminosity class III (Montes et al. 1999).
Comparison of the \ion{Na}{i} D lines with synthetic spectra shows that the line is slightly narrower than observed in the luminosity class V
thereby indicating that the luminosity class of Hen 3-1145 is IV.
We conclude that Hen 3-1145 has a spectral type M1IVe.
Fitting of the spectra with high-resolution synthetic models of a M1 star (T$_{\rm eff}$ = 3720)
shows that the ${\rm log}(g)$ that better describes the spectra is 4.0 (we used these values for
deriving the absolute magnitude of Hen~3-1145 and constrain its distance, see Sect. 3.2).

\subsection*{Hen 3-1347 = BD-18 4436}
The spectrum of Hen 3-1347 exhibits H$\delta$, H$\gamma$, H$\beta$, H$\alpha$ emission lines.
\ion{He}{ii} absorption is not present.  
\ion{He}{i} in absorption is observed at 4009,  4121, 4713, 4922, 5116, 5876, and 6678~\AA.
The absence of \ion{He}{ii} lines and the presence of \ion{He}{i} absorption lines indicate that Hen 3-1347 is a B star.
The presence of the \ion{He}{i} at 3965~\AA\  de-blended from a narrow H$\epsilon$ line,
the narrow Balmer line wings, combined with the presence of \ion{N}{ii} absorption at 3995~\AA, \ion{Si}{ii} at 4128 and 
4131~\AA, \ion{C}{ii} at 4267~\AA, and \ion{Si}{iii}  at 4553, 4568, and 4575~\AA,
indicate that Hen 3-1347 has a  luminosity class I.
A \ion{Mg}{ii} line at 4481~\AA\  slightly stronger that the \ion{He}{i} line at 4471~\AA\ 
suggest a spectral type B5 or later.
The presence of the  \ion{He}{i} line at 4009~\AA\  rules out the spectral types later than B6.
The presence of \ion{Si}{iii}  lines at 4553, 4568, and 4575~\AA, and the \ion{He}{i} line at 4121~\AA\ 
suggests  a luminosity class Ib.
The presence of the \ion{N}{ii} line at 3995~\AA\  and the mentioned \ion{Si}{iii}  lines are better described 
by the spectral type B5 than B6.
We conclude that Hen 3-1347 has a B5Ibe spectral type.

\subsection*{Hen 3-1428}
Hen 3-1428 exhibits \ion{He}{i} in absorption.
Since \ion{He}{ii} lines are absent from the spectrum, Hen 3-1428 should have a spectral type B.
Hydrogen Balmer and Paschen lines are observed in emission
displaying P-Cygni profiles.
\ion{He}{i} lines at  4471, 4713, 5016, and 5876~\AA\  exhibit P-Cygni profiles as well.
No \ion{Ca}{ii} emission is observed at 8000~\AA.
\ion{O}{i} forbidden emission is observed at 5577, 6300, and 6364~\AA.
\ion{N}{ii} emission is observed at 6548 and 6584~\AA\  showing a P-Cyni profile (the line at 5755~\AA\   is not present).
The \ion{Mg}{ii} line at 4481~\AA\ is present and is weaker than the \ion{He}{i} line at 4471~\AA\  (note, however that this
\ion{He}{i} line has PCygni profile) and is relatively broad ({\it FWHM}= 78 km/s).
The weak \ion{Mg}{ii} line at 4481 and the presence of a relatively strong \ion{He}{i} line at 4009~\AA\  indicate that 
Hen 3-1428 should be of spectral type earlier than B3.
The presence of \ion{C}{iii} lines at 4647 and 4651~\AA\  indicates that Hen 3-1428 should have a spectral type B2 or earlier.
Since in the spectrum are simultaneously present the \ion{He}{i} line at 4009~\AA,  and the \ion{O}{ii} lines at 4317 and 4320~\AA, 
the spectral type B0 is ruled out , thus the spectral type should be B1 or B2.
The presence of the \ion{N}{ii} line at 4631~\AA,  and the \ion{O}{ii} lines at 4639, 4642, and 4650~\AA\  
indicates that Hen 3-1428 is of luminosity class I or II.
As the \ion{N}{ii} at 3995~\AA\  line is stronger than the \ion{C}{ii} lines at 3919~\AA,
luminosity class of the spectrum should be class I.
The strength of the \ion{O}{i} lines at 4415 and 4417~\AA\  suggests a spectral type is B1I.
Finally, the luminosity class Ia describes better the strength of the \ion{N}{ii} 3995/\ion{He}{i} 4009 line ratio 
and the relative strength of the lines observed in the region of the \ion{N}{ii} line at 5680~\AA.
We conclude that Hen 3-1428 has a B1Iae spectral type.

\subsection*{MWC 314}
MWC 314 has a spectrum rich in emission lines of hydrogen (Balmer and Paschen series),
\ion{He}{i}, \ion{He}{ii}, \ion{Ti}{ii}, \ion{Ca}{ii}, \ion{Si}{ii}, \ion{Mg}{ii}, and \ion{Fe}{ii}.
Although the spectrum is strongly influenced by the presence of emission lines
several absorption lines are observed:
\ion{N}{ii} at 3995~\AA,
\ion{He}{i} at 4009, 4026, 4121, 4388, and 5047~\AA,
\ion{S}{ii} at 5453, 5473, and 5640~\AA,
\ion{Al}{iii} at 5696 and 5722~\AA,
and \ion{Ne}{i} at 6143, 6364 at 6462~\AA.
The absence of \ion{He}{ii} in absorption and the detection of \ion{He}{i} lines indicate that MWC 314 is a B-type star.
In particular the presence of the \ion{He}{i} lines at 4009 and 4026~\AA\  indicates that the spectral type of
MWC 314 should be early than B7 and later than B0.
The detection of the \ion{S}{ii} lines rules out the luminosity classes V and III.
The observation of the \ion{N}{ii}, \ion{Al}{iii}, and \ion{Ne}{i} lines indicates that the luminosity class is I.
The presence of the \ion{He}{i} line at 4009~\AA, the absence of the \ion{Si}{iv} lines at 4089 and 4116~\AA, 
and the \ion{C}{ii} line at 4267~\AA\  indicate that MWC 314 should have a spectral type later than B1.  
The strength of the \ion{He}{i} lines at 4009 and 5046~\AA,  and the absence of the \ion{Si}{ii} lines at 4128 and 4131~\AA\  
indicate that MWC 314 has a spectral type B3 or earlier.
The absence of the \ion{Si}{iii}  lines at 4553, 4568, and 4575~\AA\  indicates that MWC 314 should have a spectral type later than B4 if
it has luminosity class Ia and later than B3 if it has of luminosity class Ib. 
To match the \ion{Si}{ii} and \ion{Si}{iii}  requirements, MWC 314 should have a spectral type B3 and luminosity class Ib.
We conclude that MWC 314 has a spectral type B3Ib.
We note that Miroshnichenko et al. (1998) suggest a spectral type B0 for MWC 314.
This spectral type is inconsistent with our FEROS data, in particular because of the presence of the \ion{He}{i} line at 4009~\AA,
and the non detection of the \ion{Si}{iv} lines at 4089 and 4116~\AA, the \ion{C}{iii} lines at 4068, 4070, 4647, 
and 4651~\AA, the \ion{O}{ii} line at 4076~\AA,  and the \ion{Si}{iii}  lines  at 4553, 4568, and 4575~\AA.

\subsection*{MWC 593 = CD-24 13510}
MWC 593 exhibits hydrogen Balmer and Paschen emission.
The Paschen lines are broad ({\it FWHM}$\sim$500 km/s) and double peaked.
No \ion{He}{ii} is observed in absorption.
\ion{He}{i} is present, in particular the lines at 4009, 4121, and 4388~\AA.
The \ion{Mg}{ii} line at 4481~\AA\  is present, but it is very weak in contrast to the \ion{He}{i} line at 4471~\AA.
These characteristics indicate that MWC 593 is a B star of spectral types between B1 and  B5.
The lack of \ion{C}{ii} at 3919 and 3921~\AA, \ion{N}{ii} at 3995~\AA,  \ion{He}{i} at 3927~\AA,
\ion{Si}{ii} at 4128 and 4131~\AA, and \ion{Si}{iii}  at 4553, 4568, and 4575~\AA\   shows that MWC 593 is not
a giant (luminosity classes I, II, or III)
and indicates that MWC 593 should be of spectral type B3 or B4.
Comparison with spectral templates shows that the spectra that better match the relative intensity of
 the \ion{He}{i} lines and the \ion{Mg}{ii} line is the spectral type B4.
We conclude that MWC 593 has a spectral type B4Ve.

\subsection*{MWC  878 = CD-38 11837 = Hen 3-1398}
MWC 878 displays a flat spectrum rich in emission lines.
We observe in emission hydrogen Balmer lines (H$\delta$, H$\beta$, H$\alpha$),
double peaked broad ({\it FWHM}$\sim$ 390 km/s) hydrogen Paschen lines starting at 8370~\AA,
double peaked broad ({\it FWHM}$\sim$ 360 km/s) \ion{Ca}{ii} lines at  8498, 8542, and 8662~\AA,
\ion{He}{i} lines at 4009, 4471, 4713, 5876, 6678, 7066, and 7281~\AA,
\ion{He}{ii} lines at 4542, 4787~\AA,
the \ion{O}{i} line at 6300~\AA,
\ion{O}{iii} lines at 4959 and 5007~\AA,
\ion{N}{ii} lines at 5755, 6548, and 6584~\AA,
the \ion{S}{iii} line at 6312~\AA, and 
ArIII lines at 7136 and 7753~\AA.
Very few absorption lines are visible,
a narrow ({\it FWHM}$\sim$ 40 km/s) \ion{Ca}{ii}	line at 3933~\AA\ 
and the \ion{Na}{i} D lines at 5890~\AA.
We did the spectral classification based on the region around the
\ion{He}{i} lines at 4009 and  4471~\AA.
The presence of \ion{He}{i} at 4009~\AA\  in absorption indicates that the star is of  spectral type earlier
than B6.
The absence of a \ion{Mg}{ii} line at 4481~\AA\ indicates that MWC 878
has a spectral type earlier than B2 and indicates that MWC~878 is very likely not a giant or supergiant
(luminosity classes III, II, and I).
Since the \ion{Si}{iv} line at 4089~\AA, the \ion{Si}{iii}  lines at 4553, 4568, and 4575~\AA,
the \ion{C}{ii} line at 4267~\AA\ 
are absent from the spectrum, MWC 878 should have a spectral type later than B0.5.
We conclude that MWC 878 has a B1Ve spectral type.
For this star, Miroshnichenko et al. (2001) derived a spectral type of O9/B0, 
these spectral types are incompatible with our FEROS spectrum. 

\subsection*{MWC 930}
MWC 930 has a spectrum rich in emission lines.
Hydrogen Balmer and Paschen, \ion{He}{i}, strong \ion{Ca}{ii}, \ion{S}{ii}, \ion{C}{iii}, \ion{N}{ii}, \ion{Fe}{ii}, 
and [\ion{Fe}{ii}] lines are observed
displaying P-Cygni profiles.
The \ion{Na}{i} D lines at 5890~\AA\  are saturated and display a complex structure,
double absorption and three emission peaks.
The \ion{Mg}{ii} line at 4481~\AA\  is observed in absorption and it is relatively strong (EW=1.4~\AA).
Since the lines used for spectral classification are affected by emission components
the spectral classification is highly unreliable because these lines are likely to arise in an extended atmosphere
or shell.
The strong veiling also affects the spectral classification.
We attempt here to constrain the spectral classification based on the absence of strong spectral features.
Since the \ion{He}{ii} line at 4686~\AA\  is absent, MWC 930 should have a spectral type later than B0.
The detection of the \ion{He}{i} line at 5876~\AA\  implies that MWC 930 is a B-type star.
The strong \ion{Mg}{ii} line at 4481~\AA\  indicates that MWC 930 should have a spectral type later than B5.
The lack of absorption lines does not allow us to establish the luminosity class.
We conclude MWC 930 should have a spectral type B5-B9.

\subsection*{MWC 953}
MWC 953 exhibits H$\delta$ to H$\alpha$ in emission.
Its spectrum shows \ion{He}{i} in absorption and no \ion{He}{ii} lines are observed.
Since the \ion{He}{ii} line at 4686~\AA\  is absent, MWC 953 should have a spectral type later than B0.5.
The absence of the \ion{Si}{iv} line at 4089~\AA\  and the \ion{Si}{iii}  line at 4452~\AA\  indicates that 
the spectral type of MWC 953 is later than B1.
The \ion{Mg}{ii} at 4481~\AA\  line is present and is much weaker than the \ion{He}{i} line at 4471~\AA,
thus MWC 953 should have a spectral earlier than B5.
The lack of \ion{Si}{ii} lines at 4128 and 4131~\AA\ , and \ion{O}{ii} lines at 4070, 4076, 4349, 4415, and 4417~\AA\  
rules out the luminosity  classes I and II for the spectral types B2 to B5.
The presence of the \ion{C}{ii} line at 4267~\AA\  rules out the spectral types B3V, B4V and B5V.
The strength of the \ion{Mg}{ii} line at 4481~\AA\  is not compatible with the spectral types B4III and B5III.
The strength of the \ion{He}{i} line at 4713~\AA\  and the \ion{Mg}{ii} line at 4481~\AA\  is not consistent with the spectral type B3III.
The weak strength of \ion{Si}{iii}  in absorption at 4553, 4568, and 4568~\AA\  rules out the spectral type B2III.
Therefore, we  conclude that MWC 953 has a spectral type B2Ve.

\subsection*{Th 17-35}
Th 17-35 exhibits H$\alpha$ and \ion{Ca}{ii} in emission.
Its spectrum is flat and besides Balmer lines with emission components, 
very few absorption lines are observed.
No \ion{He}{ii} lines are observed.
A \ion{He}{i} line at 5976~\AA\  and a weak \ion{He}{i} line at 4471~\AA\  are present. 
In consequence, Th 17-35 is a B-type star.
The lack of strong \ion{He}{i} lines in the 4000~\AA\  region indicates that
Th 17-35 is of spectral type later than B6.
The \ion{Mg}{ii} line at 4481~\AA\  is weak but has  a strength similar to that of the \ion{He}{i} line at 4471~\AA,
suggesting that Th~17-35 should have a spectral type later than B7. 
The weak \ion{Mg}{ii} line shows that the spectral type should be earlier than B9.
Thus, Th 17-35 should have a spectral type B8.
The lack of a strong  \ion{Mg}{ii} line at 4481~\AA\  rules out the luminosity classes I and II.
The absence of the \ion{He}{i} line at 4026~\AA\  rules out the luminosity class III.
We conclude that Th~17-35 has a B8Ve spectral type.
Vieira et al. (2003) suggested a B2V spectral type for this star. 
B2 stars display the strongest \ion{He}{i} absorption lines of the B spectral class. Since we observe 
an almost flat spectrum, our data rule out the B2V type.

\subsection*{Th 35-41}
Th 35-41 does not show H$\alpha$ in emission.
The general shape of the spectrum reveals that it is a late-type star spectrum (i.e. spectral types K or M).
The narrow shape of the \ion{Na}{i} D lines at 5980~\AA\ 
indicates that Th 35-41 is a giant star, 
and suggests a spectral type later than M0III.
The lack of strong molecular bands further indicates that Th 35-41 is earlier than
M5III.
The strength of the absorption lines observed in the region close to the \ion{Na}{i} D lines
further constrains the spectral type to be earlier than M3III.
The best fit to the regions around the \ion{Na}{i} D, H$\alpha$ and \ion{Ca}{ii} ($\sim$8500~\AA) lines
(see Montes et al. 1999) was given by the spectral type M1III.
Thus, we conclude that Th 35-41 has a spectral type M1III.

\subsection*{WRAY 15-488}
WRAY 15-488 exhibits H$\alpha$ and H$\delta$
P Cyni profiles with a strong absorption component and a  weak emission component.
This suggests the presence of a wind.
\ion{He}{i} lines are not present in the spectra, thus ruling out the spectral types O and B.
Since the spectrum displays a relatively large amount of broad absorption lines,
WRAY 15-488 should have a spectral type later than A. 
The strength of the \ion{Fe}{i} and \ion{Fe}{ii} lines in the region between 4300 and 4400~\AA\  
indicates that WRAY 15-488 should have a spectral type earlier than G0.
The strength of the G-band at 4300~\AA\  indicates that WRAY~15-488 has a spectral type
earlier than F5. The G-band strength is consistent with spectral types F1-F3.
The strength of the \ion{Fe}{ii} - \ion{Ti}{ii} and \ion{Y}{ii} - \ion{Fe}{ii} doublets at 4179 and 4173~\AA,
and the presence of the \ion{Sr}{ii} lines at 4078 and 4216~\AA\  rule out the luminosity classes V and I 
(they are too strong  to be consistent with class V and too weak for being class I).
The spectrum is best matched by a spectral template F2III.
Szczerba et al. (2007) disqualified WRAY 15-488 as post-AGB star,
indicating that is a T Tauri star based on the results of Gregorio-Hetem \& Hetem (2002).
The latter authors suggest a spectral type F8, but do no specify the luminosity class.
They report the observation of H$\alpha$ in emission with an EW of -8~\AA.
The F8 spectral type is not consistent with the strength and shape of the G-band measured with 
our data. Gravity sensitive lines (e.g., \ion{Y}{ii}, \ion{Sr}{ii}) indicate that the star is not a dwarf. 
In summary, WRAY 15-488 is a giant star of spectral type F2IIIe.  
We note that in our spectra there is a marginal detection of the Li line at 6708~\AA.
 
\subsection*{WRAY 15-522}
WRAY 15-522 does not display H$\alpha$ in emission.
Its spectrum shows absorption features characteristic
of stars with spectral types late G and early K.
The strength and width of the \ion{Ca}{i} line at 4226~\AA\  indicate that 
WRAY~15-522 has a spectral type G8-K0.
The strength of the \ion{Fe}{i} line at 4271~\AA\  suggests that the spectral type
is not later than K0.
The strength and width of the \ion{Fe}{i} lines at 4532, 4787, and 5079~\AA, 
and the \ion{Mg}{I} line at 5711~\AA\ show that the spectral type is earlier than K0 but later than G8.
The strength and width of \ion{Fe}{i}+\ion{Ca}{i} lines at 5270~\AA, and the \ion{Fe}{i} lines at 5329 and 5404~\AA\  
indicate that the spectral type is earlier than K0.
Therefore, we adopt the spectral type G9 for WRAY 15-522. 
The shape and width of the \ion{Sr}{ii} lines at 4078 and 4216~\AA, 
the \ion{Y}{ii} line at 4375~\AA, and
the \ion{Ti}{ii} lines at 4400 and 4408~\AA\  (Ginestet et al. 1992, plate 44) 
indicate that the spectra is of a giant star. 
The strength of the \ion{Sr}{ii}+CN lines at 4216~\AA\  is too weak to be consistent with the luminosity 
class I and too strong to be consistent with the luminosity class V and IV. 
This indicates that WRAY 15-522 has a luminosity class III.  
We conclude that WRAY 15-522 has a spectral type G9III.

\subsection*{WRAY 15-566}
WRAY 15-566 does not display H$\alpha$ in emission.
Since its spectrum displays VO and TiO molecular bands,
WRAY 15-1566 should have a spectral type M5 or later. 
The spectrum does not display Li 6708 in absorption, thus
showing that WRAY 15-566 is not a young star.
The strength of the TiO bands at 4423, 4462~\AA\ 
indicates that WRAY 15-566 should have a spectral type M6 or earlier.
The shape and strengths of the \ion{Mn}{i} lines at 4031 and 4034~\AA,
the \ion{Ca}{i} line at 4227~\AA, the \ion{Fe}{i} line at 4326~\AA,  and the
TiO band at 4950~\AA\  show that WRAY 15-566 has a spectral type M6.
The strength of the multiple  absorption
lines inside the molecular bands at 4400 - 6000~\AA\ 
indicates that the source is a giant star of luminosity class III.
We conclude that WRAY 15-566 has a spectral type M6III.

\subsection*{WRAY 15-770}
WRAY 15-770 does not display H$\alpha$ in emission.
Its spectrum presents the TiO and VO molecular bands characteristic of stars of late-M spectral types.
The narrow sodium lines at 5890~\AA\  indicate that WRAY 15-770 is a giant star of luminosity class III.
The strength and shape of the \ion{Ca}{i} line at 4227~\AA\  indicate that WRAY 15-770
has a mid-M spectral type.
Comparison with spectral templates of the \ion{Ca}{i} line
suggests a spectral type M6 or later but earlier than M8.
The strength of the  of the TiO bands indicates that the spectral type is later than M5.
The absence of the sharp Ti I line at 4534~\AA\  indicates that WRAY 15-770
has a spectral type later than M6.
The strength and shape of the molecular bands
at  4740~\AA, 5740~\AA, 6000 - 6200~\AA\ , and 7400 - 7500~\AA\  are
best matched by the spectral type M7III.
We conclude that WRAY 15-770 has a spectral type M7III.

\subsection*{WRAY 15-1104 = CD-55 5174 = CPD-55 5588}
WRAY 15-1104 displays H$\alpha$ in emission.
\ion{He}{i} absorption lines are present in the spectrum.
The absence of the \ion{He}{ii} absorption lines, in particular the line at 4686~\AA,
indicates that the spectral type of WRAY 15-1104 is later than B0.
The presence of \ion{He}{i} absorption lines at 4009, 4121, and 4388~\AA\  shows that the
spectral type is earlier than B5.
The presence of \ion{N}{ii} line at 4631~\AA\  indicates that WRAY 15-1104 is a giant star
of luminosity class I or II and rules out the spectral types B3 to B5.
Given that the spectrum displays \ion{O}{ii} lines at 4317, 4320, 4246, 4349, 4367, 4639, and 4642~\AA,
and that the \ion{Si}{iv} line at 4089~\AA\ is present, WRAY 15--1104 should be of spectral type B1 and luminosity class I.
The almost absent \ion{C}{II} line at 4267~\AA\  and the strengths of the \ion{Si}{iv} line at 4089~\AA,
the \ion{N}{ii} line at  4631~\AA, and the \ion{O}{ii} lines at 4639 and 4642~\AA\  rule out the luminosity class Ib and 
indicate that WRAY 15-1104 should have a luminosity class Ia.
Therefore, we conclude that WRAY 15-1104 has a spectral type B1Iae.
Sarkar et al. (2005) studied WRAY 15-1104 and suggested a spectral type B1Ibe.
As explained,
our FEROS spectrum is better described by the luminosity class Ia than  the luminosity class Ib.
We note that the Sarkar et al. spectrum did not cover the wavelength range short-ward
of 4900 \AA, a region where important surface gravity indicators such as the \ion{Si}{iv} line at 4089~\AA,
the \ion{N}{ii} line at  4631~\AA, and the \ion{O}{ii} lines at 4639 and 4642~\AA\ are present. 

\subsection*{WRAY 15-1372 = CD-51 9596}
WRAY 15-1372 exhibits broad ({\it FWHM} = 382 km/s) double peaked H$\alpha$ in emission.
Its spectrum is flat and very few absorption lines are observed.
Except for the \ion{He}{i} lines at 4026 and 4471~\AA\  no other \ion{He}{i} lines are present in the spectrum.
Therefore, WRAY 15-1372 should have a spectral type B6 or later.
The strength of the \ion{Mg}{ii} line at 4481~\AA\  is similar to the strength of the \ion{He}{i} line at 4471~\AA.
Thus,  WRAY 15-1372 should have a spectral type earlier than B7.
In consequence, WRAY 15-1372 has a spectral type B6.
The absence of a strong \ion{C}{ii} line at 4267~\AA\  rules out the luminosity classes I and II.
The similar width (400 km/s) of the \ion{He}{i} line at 4471~\AA\  and the  \ion{Mg}{ii} line at 4481~\AA\ 
is not compatible with the luminosity class III.
We conclude that WRAY 15-1372 has a spectral type B6Ve.

\subsection*{WRAY 15-1435}
WRAY 15-1435 exhibits H$\alpha$, H$\beta$, \ion{Ca}{ii} at 8498 and 8662~\AA\  in emission.
Its spectrum is rich in \ion{He}{i} lines but no \ion{He}{ii} lines are observed.
Since the \ion{He}{ii} line at 4686~\AA\  is absent, WRAY 15-14135 has a spectral type later than B0.5.
The absence of the \ion{Mg}{ii} line at 4481~\AA\  shows that WRAY 15-14135 should have a spectral type earlier than B2 and  
rules out the luminosity classes I, II, and III.
These spectral characteristics indicate that WRAY 15-1435 has a spectral type B1Ve.

\subsection*{WRAY 15-1650}
The spectrum of WRAY 15-1650 displays the TiO molecular bands characteristic of M-type stars.
No H$\alpha$ emission is observed and the Li line at 6708~\AA\  is absent of the spectrum.
The strength of the molecular bands at 5000 - 8000~\AA\  indicates that WRAY 15-1650 has 
a mid-M spectral type.
The narrow \ion{Na}{i} D lines observed at 5890~\AA\  indicate that WRAY 15-1650 is a giant star
of luminosity class III (see Montes et al. 1999).
The shape of the spectrum and the strength of the molecular bands are best matched by the
M6 spectra template: the bands at 4400 and 4500~\AA\ are stronger than in the M5III template and 
the bands at 6000-7000~\AA\  are weaker than in the M7III template.
We conclude that WRAY 15-1650 has a spectral type M6III.

\subsection*{WRAY 15-1651}
WRAY 15-1651 displays H$\alpha$ and H$\beta$ emission lines.
Paschen emission is observed starting at 8438~\AA\  up to 9000~\AA.
\ion{Na}{i} D emission is observed at 5890~\AA.
The spectrum is flat and almost no absorption lines are observed.
Since the H$\alpha$ EW is of the order of 100~\AA, it is likely that this
object has a large accretion rate and that we are observing a strongly veiled spectrum.
We performed the spectral classification based on the absence of strong spectral features
that are observed in stars even with large accretion rates and strong veiling.
No molecular bands are observed; thus WRAY 15-1651 should have a spectral type earlier than M.
The \ion{Ca}{ii} line at 3933~\AA\  is not present
and the \ion{Na}{i} D lines have a {\it FWHM} (48 km/s) lower than observed in K type stars.
Therefore WRAY 15-1651 should have a spectral type earlier than K.
There is no evidence for the strong metallic lines observed in F and G type stars;
therefore WRAY 15-1561 should have a spectral type earlier than F.
The lack of \ion{Ca}{ii} and \ion{Mg}{ii} line at 4481~\AA\   in absorption suggests that WRAY 15-1561 
should have a  spectral type earlier than A.
Furthermore, the absence of the \ion{Mg}{ii} line at 4481~\AA\  indicates that WRAY 15-1561 
should have an early B spectral type.
The lack of the \ion{He}{ii} line at 4686~\AA\  suggests that WRAY 15-1561 is later than B0.
We conclude that WRAY 15-1561 should have a spectral type B1 to B5.
The lack of absorption lines does not allow us to constrain the luminosity class.

\subsection*{WRAY 15-1702}
WRAY 15-1702 has the spectrum characteristic of mid M-type stars.
It does not display H$\alpha$ in emission and the Li line at 6708~\AA\  is absent.
The narrow shape of the sodium lines at 5890~\AA\  indicates that WRAY 15-1702 is a giant star.
The shape and relative intensity of the molecular bands at 4000, 5000, 6000, and 7000~\AA\ 
indicate that WRAY 15-1702 has a spectral type later than M5.
The strength of the calcium lines at 8498 and 8662~\AA\  indicates that 
WRAY 15-1702 has a spectral type M6 or earlier.
The spectrum that best match the strength of the molecular bands and 
the 100~\AA\  region around the H$\beta$, \ion{Na}{i} D, H$\alpha$, and \ion{Ca}{ii} lines (see Montes et al. 1999)
is the spectral type M6III.
We conclude that WRAY 15-1702 has a spectral type M6III.


\begin{figure*}[ht]
\includegraphics[width=\textwidth]{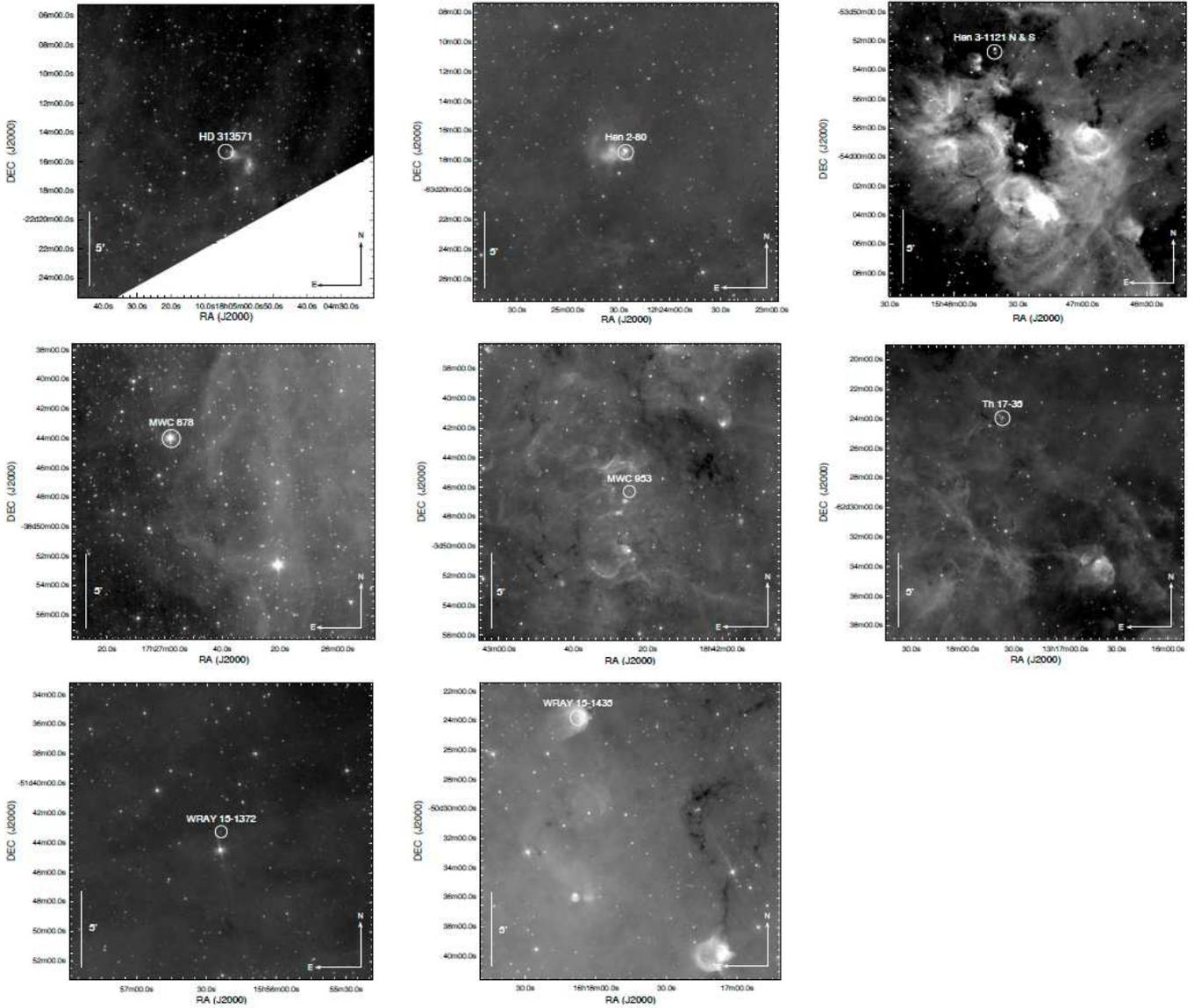}
\caption{Spitzer 8.0 $\mu$m images of the 20'$\times$20' vicinity of the confirmed Herbig Ae/Be stars not associated with
nearby star-forming regions.}
\end{figure*}

\section*{Appendix 2. IR-emission and signs of star-formation activity 
in the vicinity of  confirmed Herbig Ae/Be stars not associated with nearby SFRs.}  

We searched the Spitzer archive for 8.0 $\mu$m imaging in the 20' vicinity
of the 11 confirmed Herbig~Ae/Be stars
with distances greater than 700 pc (i.e. not associated to nearby SFRs).
Additionally, 
we queried the SIMBAD data-base for astronomical objects in the 20' vicinity 
suggesting star-formation activity such as HII regions, molecular clouds, dark clouds and young stellar objects.
Our aim was first to check whether the sources are in empty regions or
associated with large-scale IR emission and second 
to find evidence whether our sources might be members of 
distant SFRs.
We used in particular the data from the Spitzer Galactic Legacy Infrared Mid-Plane Survey Extraordinaire (GLIMPSE)\footnote{http://www.astro.wisc.edu/sirtf/} survey
and, for querying the SIMBAD data base, the Aladin tool from the Centre des Donn\'ees de Strasbourg (CDS)\footnote{http://aladin.u-strasbg.fr}.

We found Spitzer 8 $\mu$m images for 9 stars:
HD 313571, Hen~2-80, Hen~3-1121 N\&S, MWC 878, MWC~953, Th~17-35,
WRAY 15-1372, and WRAY 15-1435.
In Figure 6, we show the 20'$\times$20' vicinity of our sources.
No data were available for  Hen~3-823 and MWC~593.
We find that 7 sources, Hen~2-80, Hen~3--1121~N\&S, HD 313571, MWC~953,  WRAY~15-1435, and Th~17-35
are inside or close (separation $<5'$) to regions with extended infrared (IR) emission and 
have in their vicinity astronomical sources characteristic of SFRs. 
We suggest that these 6 Herbig Ae/Be stars are be members of distant SFRs. 
However, we note that further study of the regions is required to validate this hypothesis.
Two sources, MWC 878 and WRAY 15-1372, have no astronomical sources characteristic of SFRs and
no or very weak nearby IR emission.
They might be examples of ``isolated" Herbig Ae/Be stars.
In the following subsections, we discuss each source separately.

\subsection*{HD 313571     }
HD 313571 is embedded in a tenuous extended IR emission that increases of intensity towards the East.
In the immediate vicinity of  HD 313571 we observe at 20" and at 1,7' SW at PA$\sim$ 240$^\circ$ two bright reflection nebulae.
They are most-likely associated with the IRAS source 17580-2215. 
Palla et al. (1991) report a detection of H$_2$O maser at 22 GHz towards the position of the nebulae.
The dark nebulae LDN 237 is observed at  $\sim5$ NW (PA$\sim345^\circ$) of HD 313571.
The presence of the reflexion nebulae and the detection of masers in the vicinity of HD 313571 indicate 
that the region is young and suggest that HD 313571 may be a member of this star-forming region.

\subsection*{Hen 2-80    }
Hen 2-80 is inside an IR emission nebulosity. 
This nebula has several components:
the reflection nebulae VdBH 57b and BRAN 388, 
and at 2' SW and PA$\sim115^{\circ}$ the reflection nebulae VdBH 57d.
The presence of reflection nebulae provides further evidence of the young nature of Hen 2-80.
Given the diversity of reflection nebulae observed and the presence of several IR-Spitzer sources in the region,
we speculate that Hen 2-80 is a region where a stellar cluster is in formation. 
Further mid-IR and sub-mm imaging would be advisable to determine the number of young sources present in the region.

\subsection*{Hen 3-1121 N and Hen 3-1121 S}
Hen 3-1121 N\&S are located at the N-E of a large (20'$\times$20') 
ring-shaped IR emission region. 
A SIMBAD database search shows that this region embeds several molecular clouds 
(G328.3-0.5-46.0, G328.2-0.5-38.0, G328.2-0.5-40.3.0, G328.2-0.5-43.3, G328.2-0.5-45.8, G328.2-0.5-46.1);
OH and CH3OH masers (Caswell OH 328.237-00.547, OH 328.25-00.5, CH3OH 328.25-00.53, CH3OH 328.24-00.55,CH3OH 328.23-00.53); 
the reflection nebula GN 15.53.6; and the 
HII regions GAL 328.2-00.5,  IRAS 15541-5349 (at the center), and IRAS 15539-5353.
This ensemble of objects indicates unmistakably that this region is a region of active star formation.
Since the spectroscopy parallax distances of Hen 3-1121 S\&N  (3.1$^{+1.0}_{-0.8}$ \& 3.2$^{+6.3}_{-2.1}$ kpc respectively) 
are consistent with the kinematic distance of 3.0 kpc of the G328.236-0.547 maser (Phillips et al. 1998),
we suggest that Hen 3-1121 N\&S are members of this star-forming region. 

\subsection*{MWC 878}
West from MWC 878 we find a tenuous large scale infrared emission. 
MWC 878 is located just at the outskirts of this IR-emission.
A search in the SIMBAD database shows no sources characteristic of SFRs in the 20' vicinity.
Therefore, we suggest that MWC 878 might be an example of an ``isolated" Herbig Ae/Be star.

\subsection*{MWC 953     }
MWC 953 is embedded in an extended IR emission with a filamentary geometry. 
At $\sim$1' N and 6' S of MWC 953 we find regions strongly resembling shocks.
At 5' NW and PA$\sim290 ^\circ$ an extended dark nebula (G028.67+00.13) of size $\sim8'\times$3.5' 
is observed.
In the region several  YSOs are present (e.g., ISOGAL-P J184333.9-034459, ISOGAL-P J184329.0-034522).
The region is rich in X-ray sources.
The molecular cloud SRBY 152 is observed 5.5' SW of MWC 953 (PA$\sim216^\circ$).
The presence of these objects indicates that the region is active in star formation.
We suggest that MWC 953 might be associated to the region. Further studies of the stellar content are required to validate this conclusion.

\subsection*{Th 17-35   }
The Spitzer image of the 20' vicinity of Th 17-35 reveals a large-scale filamentary IR-emission.
The SIMBAD database shows the presence of the cloud (of unknown nature) SFO 73 at position of Th 17-35, 
the molecular clouds G306.2+0.2-42.0, G306.2+0.2-29.2, and G306.2+0.2-28.6 ($\sim$6' south),
the dark nebula DCld 306.3+00.2 (10' SE, PA$\sim$215$^{\circ}$),
and a region at 12' SW (PA$\sim$215$^{\circ}$) comprehending the reflection nebula ESO 132-4,
the cloud (of unknown nature) SFO 72, the dark cloud DCld 306.2+00.1 and the emission line star Th 17-33.
This ensemble of characteristics indicates that this is an active region of star-formation. 
Since Th 17-35 is a young star it is likely that it is a member of this region.

\subsection*{WRAY 15-1372}
WRAY 15-1372 is inside a very tenuous IR-emission.  
The SIMBAD database does not display astronomical objects associated with SFRs in the 20' vicinity of
WRAY 15-1372. 
These characteristics suggests that WRAY 15-1372 may be an ``isolated" Herbig Ae/Be star.

\subsection*{WRAY 15-1435}
WRAY 15-1435 is positionally coincident with a reflection nebula (CSI-50-16095).
All the region is embedded in a tenuous large scale IR emission.
We observe  in the region 
the  molecular cloud G332.3+0.5-96.2 approximately 6' south of WRAY 15-1435,
and the  the high-mass protostellar candidate IRAS 16082-5031 (Fontani et al. 2005) $\sim$20' SW.
At 10' SW and at PA$\sim$235$^\circ$ a dark absorption region with a shape resembling a sickle is observed.
We suggest that WRAY 15-1435 is a member of this star-formation region. However, 
further study is required to determine whether the objects observed in the region 
are physically related.


\begin{thebibliography}{}
\bibitem[Allen \& Swings(1976)]{1976A&A....47..293A} Allen, D.~A., \& Swings, J.~P.\ 1976, \aap, 47, 293 
\bibitem[S. Bagnulo, et al. 2003)]{The Messenger 2003} 
Bagnulo, S. et al., 2003, Messenger, 114, 10
\bibitem[Bertone et 
al.(2008)]{2008A&A...485..823B} Bertone, E., Buzzoni, A., Ch{\'a}vez, M., \& Rodr{\'{\i}}guez-Merino, L.~H.\ 2008, \aap, 485, 823 
\bibitem[Bohm \& Catala(1993)]{1993A&AS..101..629B} B\"ohm, T., \& Catala, C.\ 1993, \aaps, 101, 629 
\bibitem[Brown et al.(2007)]{2007ApJ...664L.107B} Brown, J.~M., et al.\ 
2007, \apjl, 664, L107 
\bibitem[Calvet et al.(2005)]{2005ApJ...630L.185C} Calvet, N., et al.\ 
2005, \apjl, 630, L185 
bibitem[Carmona]{Carmona 2010} Carmona, A. 2010, in proceedings of the conference Origin and Evolution of Planets 2008, Ascona, Switzerland.
Eds. L. Mayer. Earth, Moon, Planets, in press. astroph/2009arXiv0911.2271C.
\bibitem[Dong \& Hu(1991)]{1991AcApS..11..172D} Dong, Y.-S., \& Hu, J.-Y.\ 1991, Acta Astrophysica Sinica, 11, 172 
\bibitem[Finkenzeller 
\& Mundt(1984)]{1984A&AS...55..109F} Finkenzeller, U., \& Mundt, R.\ 1984, \aaps, 55, 109 
\bibitem[Fontani et 
al.(2005)]{2005A&A...432..921F} Fontani, F., Beltr{\'a}n, M.~T., Brand, J., Cesaroni, R., Testi, L., Molinari, S., \& Walmsley, C.~M.\ 2005, \aap, 432, 921 
\bibitem[Fukagawa et al.(2004)]{2004ApJ...605L..53F} Fukagawa, M., et al.\ 
2004, \apjl, 605, L5
\bibitem[Ginestet 1992]{Ginestet 1992} Ginestet, N., Carquillat, J.M., Jaschek, M., Jaschek, C., 
Pedoussaut, A., Rochette, J., Atlas de spectres stellaires de standards de classification MK, binaires spectroscopiques, \'etoiles particuli\'eres. Observatoire Midi-Pyr\'en\'ees, Toulouse CRDP 1992. 
\bibitem[Grady et al.(2005)]{2005ApJ...630..958G} Grady, C.~A., et al.\ 
2005, \apj, 630, 958 
\bibitem[Gregorio-Hetem \& Hetem(2002)]{2002MNRAS.336..197G} Gregorio-Hetem, J., \& Hetem, A.\ 2002, \mnras, 336, 197 
\bibitem[Guimar{\~a}es et 
al.(2006)]{2006A&A...457..581G} Guimar{\~a}es, M.~M., Alencar, S.~H.~P., Corradi, W.~J.~B., \& Vieira, S.~L.~A.\ 2006, \aap, 457, 581
\bibitem[Hartmann(1999)]{1999NewAR..43....1H} Hartmann, L.\ 1999, New 
Astronomy Review, 43, 1 
\bibitem[Herbig(1960)]{1960ApJS....4..337H} Herbig, G.~H.\ 1960, \apjs, 4, 
337 
\bibitem[Hilton 
\& Lahulla(1995)]{1995A&AS..113..325H} Hilton, J., \& Lahulla, J.~F.\ 1995, \aaps, 113, 325
\bibitem[Kaufer, A. et al. 1999]{1999Messenger...95..8} Kaufer, A. et al. 1999, The Messenger 95, 8. 
\bibitem[Lang 1991]{Lang, K.R.}  Lang, K.R.., Astrophysical Data: Planets and Stars., Springer, 1991
\bibitem[Lanz \& Hubeny(2007)]{2007ApJS..169...83L} Lanz, T., \& Hubeny, I.\ 2007, \apjs, 169, 83 
\bibitem[Mannings \& Sargent(1997)]{1997ApJ...490..792M} Mannings, V., \& 
Sargent, A.~I.\ 1997, \apj, 490, 792 
\bibitem[Mer{\'{\i}}n et al.(2008)]{2008ApJS..177..551M} Mer{\'{\i}}n, B., 
et al.\ 2008, \apjs, 177, 551 
\bibitem[Mikami \& Heck(1982)]{1982PASJ...34..529M} Mikami, T., \& Heck, A.\ 1982, \pasj, 34, 529
\bibitem[Miroshnichenko et al.(1998)]{1998A&AS..131..469M} Miroshnichenko, A.~S., Fremat, Y., Houziaux, L., Andrillat, Y., Chentsov, E.~L., \& Klochkova, V.~G.\ 1998, \aaps, 131, 469 
\bibitem[Miroshnichenko et 
al.(2001)]{2001A&A...371..600M} Miroshnichenko, A.~S., Levato, H., Bjorkman, K.~S., \& Grosso, M.\ 2001, \aap, 371, 600
\bibitem[Miroshnichenko et al.(2005)]{2005A&A...436..653M} Miroshnichenko, A.~S., Bjorkman, K.~S., Grosso, M., Hinkle, K., Levato, H., \& Marang, F.\ 2005, \aap, 436, 653 
\bibitem[Montes et al.(1999)]{1999ApJS..123..283M} Montes, D., Ramsey, 
L.~W., \& Welty, A.~D.\ 1999, \apjs, 123, 283 
\bibitem[Mottram et al.(2007)]{2007MNRAS.377.1363M} Mottram, J.~C., Vink, 
J.~S., Oudmaijer, R.~D., \& Patel, M.\ 2007, \mnras, 377, 1363 
\bibitem[Morgan et al.(1943)]{1943QB881.M6.......} Morgan, W.~W., Keenan, 
P.~C., \& Kellman, E.\ 1943, Chicago, Ill., The University of Chicago press [1943]
\bibitem[Muzerolle et al.(2004)]{2004ApJ...617..406M} Muzerolle, J., 
D'Alessio, P., Calvet, N., \& Hartmann, L.\ 2004, \apj, 617, 406 
\bibitem[Najita et al.(2007)]{2007prpl.conf..507N} Najita, J.~R., Carr, 
J.~S., Glassgold, A.~E., \& Valenti, J.~A.\ 2007, in Protostars and Planets V, 507
\bibitem[Palla et 
al.(1991)]{1991A&A...246..249P} Palla, F., Brand, J., Comoretto, G., Felli, M., \& Cesaroni, R.\ 1991, \aap, 246, 249 
\bibitem[Phillips et al.(1998)]{1998MNRAS.300.1131P} Phillips, C.~J., 
Norris, R.~P., Ellingsen, S.~P., 
\& McCulloch, P.~M.\ 1998, \mnras, 300, 1131
\bibitem[Pontefract et al.(2000)]{2000MNRAS.319L..19P} Pontefract, M., 
Drew, J.~E., Harries, T.~J., \& Oudmaijer, R.~D.\ 2000, \mnras, 319, L19 
\bibitem[Preibisch 
\& Mamajek(2008)]{2008hsf2.book..235P} Preibisch, T., \& Mamajek, E.\ 2008, Handbook of Star Forming Regions, Volume II, 235
\bibitem[Reipurth et 
al.(1996)]{1996A&AS..120..229R} Reipurth, B., Pedrosa, A., \& Lago, M.~T.~V.~T.\ 1996, \aaps, 120, 229
\bibitem[Sarkar et 
al.(2005)]{2005A&A...431.1007S} Sarkar, G., Parthasarathy, M., \& Reddy, B.~E.\ 2005, \aap, 431, 1007  
\bibitem[Schmidt-Kaler 1982]{Schmidt-Kaler 1982}
Schmidt-Kaler, Th. 1982, in Landolt/Bornstein, New Series Group VI, Vol. 2
\bibitem[Schonberner 
\& Drilling(1984)]{1984ApJ...278..702S} Sch\"onberner, D., \& Drilling, J.~S.\ 1984, \apj, 278, 70
\bibitem[Schultz \& Wiemer(1975)]{1975A&A....43..133S} Schultz, G.~V., \& Wiemer, W.\ 1975, \aap, 43, 133 
\bibitem[Semenov et al.(2005)]{2005ApJ...621..853S} Semenov, D., 
Pavlyuchenkov, Y., Schreyer, K., Henning, T., Dullemond, C., 
\& Bacmann, A.\ 2005, \apj, 621, 853 
\bibitem[Soubiran et al.(1998)]{1998yCat..41330221S} Soubiran, C., Katz, 
D., \& Cayrel, R.\ 1998, VizieR Online Data Catalog, 413, 30221 
\bibitem[Su{\'a}rez et 
al.(2006)]{2006A&A...458..173S} Su{\'a}rez, O., Garc{\'{\i}}a-Lario, P., Manchado, A., Manteiga, M., Ulla, A., \& Pottasch, S.~R.\ 2006, \aap, 458, 173 
\bibitem[Szczerba et 
al.(2007)]{2007A&A...469..799S} Szczerba, R., Si{\'o}dmiak, N., Stasi{\'n}ska, G., \& Borkowski, J.\ 2007, \aap, 469, 799 
\bibitem[The et 
al.(1994)]{1994A&AS..104..315T} Th\'{e}, P.~S., de Winter, D., \& Perez, M.~R.\ 1994, \aaps, 104, 315 
\bibitem[van den Bergh(1966)]{1966AJ.....71..990V} van den Bergh, S.\ 1966, 
\aj, 71, 990
\bibitem[Vieira et al.(2003)]{2003AJ....126.2971V} Vieira, S.~L.~A., 
Corradi, W.~J.~B., Alencar, S.~H.~P., Mendes, L.~T.~S., Torres, C.~A.~O., 
Quast, G.~R., Guimar{\~a}es, M.~M., \& da Silva, L.\ 2003, \aj, 126, 2971 
\bibitem[Vink et 
al.(2005)]{2005A&A...430..213V} Vink, J.~S., Harries, T.~J., \& Drew, J.~E.\ 2005, \aap, 430, 213 
\bibitem[Vink et al.(2002)]{2002MNRAS.337..356V} Vink, J.~S., Drew, J.~E., 
Harries, T.~J., \& Oudmaijer, R.~D.\ 2002, \mnras, 337, 356 
\bibitem[Waters \& Waelkens(1998)]{1998ARA&A..36..233W} Waters, L.~B.~F.~M., \& Waelkens, C.\ 1998, \araa, 36, 233 
\bibitem[de Zeeuw et al.(1999)]{1999AJ....117..354D} de Zeeuw, P.~T., 
Hoogerwerf, R., de Bruijne, J.~H.~J., Brown, A.~G.~A., 
\& Blaauw, A.\ 1999, \aj, 117, 354 
\end{thebibliography}
\end{document}